\documentclass[3p,11pt,review]{elsarticle}

\usepackage{adjustbox}
\usepackage{booktabs}
\usepackage{xcolor}
\usepackage{accents}

\usepackage[markup=underlined]{changes}
\usepackage{color}

\usepackage{lineno,hyperref}

\journal{Computers and Fluids}

\begin{document}

\begin{frontmatter}

\title{Open-source finite volume solvers for multiphase (n-phase) flows involving either Newtonian or non-Newtonian complex fluids}

\author{Chris Dritselis}
\ead{dritselis@auth.gr}

\author{George Karapetsas\corref{mycorrespondingauthor}}
\cortext[mycorrespondingauthor]{Corresponding author}
\ead{gkarapetsas@auth.gr}

\address{Department of Chemical Engineering, Aristotle University of Thessaloniki, \\Thessaloniki 54124, Greece}

\begin{abstract}
\linespread{1.1}
\small {Two new control volume solvers 
\texttt{multiFluidInterFoam} and \texttt{rheoMultiFluidInterFoam} are presented for the simulation of Newtonian and non-Newtonian n-phase flows ($n \geq 2$), respectively, fully accounting for interfacial tension and contact-angle effects for each phase. The \texttt{multiFluidInterFoam} solver modifies certain crucial aspects of the regular \texttt{multiphaseInterFoam} solver provided by \texttt{OpenFOAM} for Newtonian flows, improving its efficiency and robustness, but most importantly improving considerably its accuracy for surface tension driven flows. The new solver uses the volume-of-fluid (VOF) method based on the MULES method and artificial interface compression for the interface capturing, in combination with a suitable continuum surface force model; i.e. the interfacial tension coefficient decomposition method is employed to treat pairings of interfacial tension between the phases. VOF smoothing is also implemented by applying a Laplacian filter on the distribution of volume fractions, and the \texttt{SIMPLEC} algorithm is employed for the velocity-pressure coupling. The above algorithm has also been extended with the development of the \texttt{rheoMultiFluidInterFoam} solver which is capable of fully taking into account complex non-Newtonian effects (e.g. yield stress, viscoelasticity, etc.) of the involved liquids. To this end, the developed solver incorporates the \texttt{RheoTool} toolbox (Pimenta and Alves (2017) J. Non-Newtonian Fluid Mech. 239, 85--104) \cite{Pimenta2017}, utilizing a wealth of constitutive equations suitable for modelling different types of fluids with complex rheology. Our solvers are tested for a wide range of different flow setups, considering typical benchmark two-phase and three-phase flows, such as the cases of dam break with an obstacle, spreading of a floating lens, a levitating drop, and a bubble rising in an ambient fluid; flows involving Newtonian, viscoplastic and viscoelastic liquids have been considered. Comparisons against analytical solutions or previously published numerical data are performed clearly demonstrating the capability of the new solvers to efficiently and accurately simulate interfacial tension dominated three-phase flows (and n-phase flows in general) for both simple and complex liquids.}
\end{abstract}

\begin{keyword}
\linespread{1.2}
\small{
multiphase flow \sep volume-of-fluid \sep contact line \sep non-newtonian \sep floating lens \sep drop levitation \sep \texttt{multiFluidInterFoam} \sep \texttt{rheoMultiFluidInterFoam}}
\end{keyword}

\end{frontmatter}


\section{Introduction}\label{sec1}

Multiphase flows, where two or more fluids have interfacial surfaces, frequently arise in many industrial engineering and scientific applications. A typical example is oil recovery, which involves at least three-phase fluids including oil, water, and air. Other examples are nuclear reactors and microfluidics. The flow in all these applications exhibits a wide range of spatial and temporal scales, and it includes various complex interactions between the fluids, the motion of droplets and bubbles, as well as the formation of films and jets. Despite the fact that there is a sufficient number of numerical studies on two-phase flows, research on cases where a higher number of phases may be involved, e.g. three-phase flows, is still very limited, partly because of the variety and complexity of interactions between the phases and difficulties in the numerical treatment of special locations where the three interfaces connect, i.e., triple junction points. With the rapid development of computer technology, numerical modeling provides reliable and accurate tools and it has a considerable role in the detailed investigation of multiphase flows. In the past, many efforts have been made to develop numerical approaches for the simulation of three-phase flows based on the numerical solution of Navier--Stokes equations together with various approaches devoted to the capturing or tracking of the interfaces between fluids, such as the level-set \cite{Zhao1996, Smith2002}, front-tracking \cite{Muradoglu2010}, volume-of-fluid \cite{Bonhomme2012, Xie2020}, smoothed particle hydrodynamics \cite{Tofighi2013}, and phase field \cite{Garcke1999, Kim2004, Kim2007} methods. Usually, the fluids under consideration are Newtonian, with far less attention being paid to the use and development of such methods in the study of three-phase flows when one or more fluids exhibits a non-Newtonian behavior.

The present study is focused on the development of efficient and accurate numerical tools for the investigation of interfacial tension dominated flows involving more than two phases, such as for example, the three-phase flow of a droplet spreading between two stratified fluids. Earlier attempts to address the modelling of $n$-phase Newtonian flows under the framework of \texttt{OpenFOAM} \cite{Weller1998, OpenFOAM} have culminated in the development of the \texttt{multiphaseInterFoam} solver \cite{Andersson2010}.

As it will be discussed in detail below, the latter solver performs quite well in inertia dominated flows, but exhibits very poor performance in flows dominated by surface tension. To address this important issue, a new generic multiphase solver is presented, named \texttt{multiFluidInterFoam}, capable to accurately model Newtonian $n$-phase flows, $n \geq 2$, and fully taking into account interfacial phenomena and contact line effects. The new solver has been developed building upon and considerably improving, the regular \texttt{multiphaseInterFoam} solver provided by \texttt{OpenFOAM} \cite{Weller1998, OpenFOAM}. The numerical framework of the developed solver consists of a finite volume method and the volume-of-fluid method (VOF) that employs a volumetric phase fraction for each fluid phase. Sharpening of the interface is achieved by introducing a properly adapted artificial compression term in the equation for each volume fraction. Smoothing of the volume fraction of each phase is also suitably incorporated in the calculation of curvature of interfaces for $n$-phase flows. The multiphase nature of the system is taken into account by a continuum surface force model. In contrast to the original multiphase solver, the interfacial tension coefficient decomposition method is used here to handle the pairings of interface tension between the different phases. 

The \texttt{multiFluidInterFoam} solver has been developed for Newtonian fluids. However, a situation where one or more fluids exhibits a non-Newtonian behavior is not unusual. Such flows may arise in a number of different setups ranging from environmental applications (e.g. spreading of weathered oil spills exhibiting complex rheology due to evaporation of the most volatile components) to biomedical applications (e.g., inhaled drug in the form of aerosol droplets spreading along the mucus lining of the lung, which commonly exhibits viscoelastic properties and yield stress) \cite{Khanal2015,Sharma2015}.

To address the modelling of such flows, a second solver \texttt{rheoMultiFluidInterFoam} is developed as an extension of the aforementioned \texttt{multiFluidInterFoam}, able to handle multiphase flows, where any number of the involved phases can exhibit a rheologically complex behaviour. To the best of our knowledge, the developed solver presented herein is the first fully capable solver of handling and accurately solving surface tension driven $n$-phase flows involving any number of rheologically complex fluids, including viscoelastic or elasto-viscoplastic effects. This goal has been achieved by combining the VOF method, properly adapted for multiple fluids, with the \texttt{RheoTool} toolbox \cite{rheoTool2016}, in order to make use of several constitutive equations for the simulation of generalized Newtonian and viscoelastic fluids. 

The new multiphase solvers are both validated through comparisons with existing analytical and numerical solutions in the case of Newtonian fluids. The problems of the dam break, a floating lens, and the levitation of drop initially located within two fluids are used here for comparison purposes. Since there is no rigorous benchmark for non-Newtonian three-phase fluid flows, for the validation of the \texttt{rheoMultiFluidInterFoam} solver the following approach is adopted. First, we examine the two-phase problem of a bubble rising both in a viscoplastic (described by Herschel--Bulkley model) and a viscoelastic (described by Phan-Thien Tanner model) fluid, as a reference to validate the presently \texttt{RheoTool} based implementation of the model. Finally, consistency verification tests are performed for the floating lens and levitating drop problems, which are re-examined by assuming that the bottom fluid layer exhibits a viscoplastic behavior.

The remainder of this article is organized as follows. Description of the models and numerical methods is given in Section \ref{sec2}. Results are shown and compared with analytical solutions or recently published numerical data in Section \ref{sec3}. Finally, a discussion is presented in Section \ref{sec4} and some concluding remarks are summarized in Section \ref{sec5}. In addition to describe and provide access to the source codes of a robust open-source three-phase flow solvers (available for download on GitHub: \url {https://github.com/gkarapetsas/multiFluidInterFoam} and \url {https://github.com/gkarapetsas/rheoMultiFluidInterFoam}).

\section{Governing equations and numerical methods}\label{sec2}

We begin by presenting in summary the main features of the standard \texttt{multiphaseInterFoam} solver provided by \texttt{OpenFOAM} and then proceed with a detailed presentation of the key features of the newly developed solvers.

\subsection{Overview of the original \texttt{multiphaseInterFoam} solver provided by \texttt{OpenFOAM}} \label{sec2.1}

The standard multiphase solver available in \texttt{OpenFOAM}-v7 is \texttt{multiphaseInterFoam}. In contrast to the two-phase incompressible flow solver \texttt{interFoam}, it accounts for several incompressible fluids with interface capturing, surface-tension and contact-angle effects for each phase. Several non-Newtonian models are available at runtime, including a bounded bi-viscosity version of the Herschel--Bulkley model, which will also be used in the present work for comparison purposes.

In multi-component flows, several components are present in one or more phases. The volume fraction of component $i$ is denoted as $\alpha_i$, where $i = 1, 2, ..., N_c$, and $N_c$ is the number of components. In the present work, we shall focus on interfacial flows with three components and, thus, consider $N_c=3$. The density and dynamic (kinematic) viscosity of component $i$ are $\rho_i^\ast$ and $\mu_i^\ast \, (\nu_i^\ast)$, respectively; the
asterisk indicates a dimensional quantity. The governing equations for multiphase flows are the equation of continuity, momentum and of transport of the volume fraction distributions. The continuity equation is 

\begin{equation} \label{eqA2}
\nabla\cdot \mathbf{u}^\ast = 0 \, .
\end{equation}

\noindent
The incompressible Navier--Stokes equation (NSE) is reformulated in order to take into account the multiphase nature of the system as

\begin{equation} \label{eqA1}
\frac{ \partial \rho^\ast \mathbf{u}^\ast }  {\partial t^\ast} +\nabla\cdot(\rho^\ast \mathbf{u}^\ast \mathbf{u}^\ast) = -\nabla p_d^\ast + \mathbf{g}^\ast \cdot \mathbf{x}^\ast \nabla\rho^\ast + \nabla \cdot \mathbf{\underline{\tau}}^\ast + \mathbf{f}_{st}^\ast \, .
\end{equation}

\noindent
In the above equations, $\textbf{u}^\ast$ is the velocity, $t^\ast$ is the time, $\mathbf{g}^\ast$ is the gravity acceleration, $\mathbf{x}^\ast$ is the position vector, $\mathbf{\underline{\tau}}^\ast$ is the deviatoric stress tensor, $\rho^\ast$ denotes the bulk fluid density and $\textbf{f}_{st}^\ast$ is the source of momentum due to the surface tension. The quantity $p_d^\ast$ is a modified pressure containing the static pressure $p^\ast$ and the hydrostatic component $p_d^\ast=p^\ast-\rho^\ast \mathbf{g}^\ast \cdot \mathbf{x}^\ast$. 

In general, the fluids can be considered generalized Newtonian with
\begin{equation}\label{eqA3b}
\mathbf{\underline{\tau}}^\ast = \rho^\ast \nu^\ast \left(  \dot{\gamma}^\ast \right) \mathbf{\dot{\underline{\gamma}}^\ast},
\end{equation}

\noindent
where $\mathbf{\dot{\underline{\gamma}}^\ast} = \nabla \mathbf{u}^\ast + \nabla {\mathbf{u}^\ast}^\mathrm{T}$ is the rate of strain tensor,  $\dot{\gamma}^\ast=\sqrt{(\mathbf{\dot{\underline{\gamma}}}^\ast:{\mathbf{\dot{\underline{\gamma}}}^\ast)/2}}$ is its second invariant and $\nu^\ast$ is the bulk kinematic viscosity.

The capturing of interfaces is performed by the volume of fluid method (VOF) \cite{Hirt1981} utilizing the volumetric phase fraction $\alpha_i$ for each fluid $i$. The transport equation for the $i$th volume fraction $\alpha_i$ is

\begin{equation} \label{eqA4}
\frac{ \partial \alpha_i}  {\partial t^\ast} +\nabla\cdot(\alpha_i \mathbf{u}_i^\ast) = 0, \, \,\,\,\,\,\, \,\,\,\,\,\, \,\,\,\,\, i=1,2,...N_c \, ,
\end{equation}

\noindent 
where $\mathbf{u}_i^\ast$ is the velocity field corresponding to the $i$th fluid. The volume fractions $\alpha_i$ are coupled with each other by the normalization:

\begin{equation} \label{eqA5}
\sum_{i=1}^{N_c} \alpha_i = 1, \, \,\,\,\,\,\, \,\,\,\,\,\, \,\,\,\,\, \alpha_i \in [0,1] \, .
\end{equation}

\noindent
The bulk density $\rho^\ast$ and the bulk viscosities $\mu^\ast, \, \nu^\ast$ are treated as volumetric mixture values and they are computed in each computational cell using the densities $\rho_i^\ast$ and viscosities $\mu_i^\ast, \, \nu_i^\ast$ of each fluid and the volumetric phase fractions as weighting factors as

\begin{equation} \label{eqA6}
\rho^\ast=\sum_{i=1}^{N_c} \alpha_i \rho_i^\ast \, ,
\end{equation}

\begin{equation} \label{eqA8}
\nu^\ast \left(\dot{\gamma}^\ast \right) =\frac{1}{\rho^\ast}\sum_{i=1}^{N_c} \alpha_i \rho_i^\ast \nu_i ^\ast \left(\dot{\gamma}^\ast \right) \, .
\end{equation}

In the case of a purely Newtonian fluid, the viscosity is simply considered to be constant

\begin{equation} \label{eqA3a}
\nu_i^\ast\left(\dot{\gamma}^\ast \right) = \nu_{i,Newtonian}^\ast = \mbox{constant} \, ,
\end{equation}

\noindent
whereas other available constitutive equations such as the power-law, Carrreau and Herschel--Bulkley models are available to the user. For the purposes of the present study we will consider the case of a  viscoplastic material described by the Herschel--Bulkley model

\begin{equation} \label{eqA3}
\nu_i^\ast\left(\dot{\gamma}^\ast \right) = \min\left( \nu_{0,i}^\ast, \tau_{0,i}^\ast/\dot{\gamma}^\ast + k_i^\ast(\dot{\gamma}^\ast)^{n_i-1}  \right) \, ,
\end{equation}

\noindent
which combines the effects of Bingham plastic and power-law behavior in a fluid. For low strain rates $\dot{\gamma}^\ast$, the material is modeled as a very viscous fluid with viscosity $\nu_{0,i}^\ast$. Beyond a threshold in strain-rate corresponding to a threshold stress $\tau_{0,i}^\ast$, the viscosity is described by a power law; where $k_i^\ast$ is the flow consistency index and $n_i$ is the power-law index of each phase. 

The effect of surface tension is implemented using the Continuum-Surface-Force (CSF) model of Brackbill et al. \cite{Brackbill1992} as a volumetric force near the interface as \cite{Personnettaz2018, Bublik2020}

\begin{equation} \label{eqA9}
\mathbf{f}_{st}^\ast = \sum_i \sum_{j \neq i} \sigma_{ij}^\ast \kappa_{ij}^\ast \delta_{ij}^\ast\, ,
\end{equation}

\noindent
with $\sigma_{ij}^\ast$ denoting the constant interface tension between phases $i$ and $j$. In this approach the interface is neither tracked explicitly nor shape or location are known and, thus, an exact boundary condition cannot be applied at the interfaces. The curvature of the interface $ij$ is

\begin{equation} \label{eqA10}
\kappa_{ij}^\ast= -\nabla \cdot \frac{\alpha_j\nabla\alpha_i-\alpha_i\nabla \alpha_j}{\mid\alpha_j\nabla\alpha_i-\alpha_i\nabla \alpha_j\mid \,+\, sf} \, ,
\end{equation}

\noindent
where $sf$ is a stabilization factor that accounts for non-uniformity of the computational mesh $sf=\epsilon/\left( \sum_{i=1}^{\mathrm{ncv}} V_i/\mathrm{ncv}\right)^{(1/3)}$, $\mathrm{ncv}$ is the number of computational cells, $V_i$ is the volume of cell $i$, and $\epsilon$ is a small positive number, typically $\epsilon=10^{-8}$. The term $\delta_{ij}^\ast = \alpha_j\nabla\alpha_i-\alpha_i\nabla \alpha_j$ applies the volumetric force only near the interfaces, where variations of the indicator function $\delta_{ij}^\ast$ are present. Thus, the surface tension force becomes

\begin{equation} \label{eqA11}
\mathbf{f}_{st}^\ast=-\sum_i \sum_{j \neq i} \sigma_{ij}^\ast \nabla \cdot \left( \frac{\alpha_j\nabla\alpha_i-\alpha_i\nabla \alpha_j}{\mid\alpha_j\nabla\alpha_i-\alpha_i\nabla \alpha_j\mid \,+\, sf}\right)  (\alpha_j\nabla\alpha_i-\alpha_i\nabla \alpha_j) \, .
\end{equation}

The fluid interfaces are sharpened by introducing artificial compression terms into Eq. (\ref{eqA4}), which it may be re-written as

\begin{equation}
\frac{\partial a_i} {\partial t^\ast} + \nabla \cdot \left( a_i \mathbf{u}^\ast + \sum_{k \neq i, k=1}^{N_c}  a_i a_k \mathbf{u_r}_{,ik}^\ast \right)=0, \,\,\,\,\,\,\,\,\,\,\,\,  i=1,2,...,N_c \, ,
\label{VOFComp1}
\end{equation}

\noindent
where $\mathbf{u_r}_{,ik}^\ast$ is the artificial compression velocity given by 

\begin{equation}
(\mathbf{u_r}_{,ik}^\ast)_\mathrm{f} = (\mathbf{n}_{ik})_\mathrm{f} \min \left[ C_{a_i} \frac{|\phi_{\mathrm{f}}^\ast|}{|\mathbf{S}_\mathrm{f}^\ast|}, \max \left( \frac{|\phi_{\mathrm{f}}^\ast|}{|\mathbf{S}_\mathrm{f}^\ast|}\right) \right], \,\,\,\,\,\,\,\,\,\,\,\,  i=1, 2,...,N_c \,\,\, \mathrm{and} \,\,\, k \ne i \, ,
\label{VOFComp2}
\end{equation}

\noindent
where $(\mathbf{n}_{ik})_\mathrm{f}$ is the normal vector of the cell surface, $\phi_{\mathrm{f}}^\ast$ is the volume flux given as $\phi_{\mathrm{f}}^\ast=\mathbf{u}_\mathrm{f}^\ast \cdot \mathbf{S}_\mathrm{f}^\ast$, $\mathbf{S}_\mathrm{f}^\ast$ is the cell surface area, and $C_{a_i}$ is an adjustable coefficient, the value of which can be set between $0$ and $4$. The derivation of Eq. (\ref{VOFComp1}) is shown in the Appendix.
The quantity $(\mathbf{u_r}_{,ik}^\ast)_\mathrm{f}$ can be interpreted as a relative velocity between fluids $i$ and $k$, which arises from the changes in the density and viscosity within the fluid interface. By taking the divergence of the compression velocity $\mathbf{u_r}_{,ik}^\ast$, the conservation of the VOF function is guaranteed \cite{Weller2008}. The term $a_i\,a_k$ in Eq. (\ref{VOFComp1}) ensures that the compression term is active only in the interfacial area where $0 < a_i < 1$. The level of compression depends on the value of $C_{a_i}$. There is no compression for $C_{a_i}=0$, a moderate compression exists for $0 < C_{a_i} \le 1$, while an enhanced compression takes place for $1 < C_{a_i} \le 4$ \cite{Berberovic2009, Deshpande2012}. The normal vector $(\mathbf{n}_{ik})_\mathrm{f}$ is calculated as 

\begin{equation} \label{VOFComp3}
(\mathbf{n}_{ik})_\mathrm{f} = \frac{ (\alpha_{k})_\mathrm{f} (\nabla \alpha_{i})_\mathrm{f} - (\alpha_{i})_\mathrm{f} (\nabla \alpha_{k})_\mathrm{f} }  {\mid (\alpha_{k})_\mathrm{f} (\nabla \alpha_{i})_\mathrm{f} - (\alpha_{i})_\mathrm{f} (\nabla \alpha_{k})_\mathrm{f} \mid \, + \, sf} \, .
\end{equation}

\noindent
In the \texttt{multiphaseInterFoam} solver, the governing equations are solved sequentially in a segregated manner, where the transport equations for the volume fractions (Eqs. \ref{eqA4}, \ref{eqA5}) are first solved, the properties of the mixture (Eqs. \ref{eqA6}-\ref{eqA8}) are then calculated based on the the constitutive equations (Eqs. \ref{eqA3a}, \ref{eqA3}), followed by the solution of the momentum equation (Eq. \ref{eqA1}) and the continuity (pressure) equation (Eq. \ref{eqA2}). The velocity-pressure coupling is ensured using the \texttt{PIMPLE} algorithm that is a combination of \texttt{PISO} \cite{Issa1986} (Pressure-Implicit Split Operator) and \texttt{SIMPLE} \cite{Patankar1972} (Semi-Implicit Method for the Pressure-Linked Equation) algorithms.

\subsection{Overview of the \texttt{multiFluidInterFoam} solver}\label{sec2.2}

In the following sections, we describe the new \texttt{multiFluidInterFoam} solver and the main modifications introduced in the original multiphase solver, which are suitable for the accurate simulation of interfacial tension dominated, three-phase flows; the method should in principle be valid for any n-phase flows with $n\geq2$. These changes can considerably increase the stability and computational speed of the new solver, while keeping its time and space accuracy.

\subsubsection{Interfacial tension model} \label{sec2.2.1}
The interfacial tension force in two-phase flows is usually obtained by using the Continuum-Surface-Force (CSF) \cite{Brackbill1992} method as $\mathbf{f}_{st}^\ast=\sigma_{12}^\ast \kappa_{12}^\ast \mathbf{n}_{12} \delta^\ast$, where $\sigma_{12}^\ast$ is the interfacial tension coefficient between fluid $1$ and $2$, $\kappa_{12}^\ast =\nabla\cdot \mathbf{n}_{12}$ is the interfacial curvature, $\mathbf{n}_{12}$ is the interface unit normal, and $\delta^\ast$ is the Dirac delta function. In three-phase flows, the fluid flow might be influenced simultaneously by more than two interfaces adjacent to triple junction points, and, thus, the above definition has to be further extended. 

As shown in Section \ref{sec2.1}, the original \texttt{multiphaseInterFoam} solver adopts Eq. (\ref{eqA11}) in order to account for the surface tension. In the present work, the interfacial tension coefficient decomposition method is employed to deal with tension pairings between different phases using a compositional approach \cite{Smith2002, Xie2020, Tofighi2013, Kim2007}. The physical interfacial tension coefficients $\sigma_{ij}^\ast$ between phase $i$ and phase $j$ are decomposed into phase-specific interfacial tension coefficients as $\sigma_{ij}^\ast = \Sigma_i^\ast + \Sigma_j^\ast$. In particular, for a three-phase flow ($N_c=3$) once the physical interfacial tension coefficients $\sigma_{ij}^\ast$ are known, the phase-specific interfacial tension coefficients can be obtained as

\begin{equation}\label{eqB1}
\Sigma_1^\ast=0.5(\sigma_{12}^\ast+\sigma_{13}^\ast-\sigma_{23}^\ast),
\end{equation}
\begin{equation}\label{eqB2}
\Sigma_2^\ast=0.5(\sigma_{12}^\ast+\sigma_{23}^\ast-\sigma_{13}^\ast),
\end{equation}
\begin{equation}\label{eqB3}
\Sigma_3^\ast=0.5(\sigma_{13}^\ast+\sigma_{23}^\ast-\sigma_{12}^\ast).
\end{equation}

\noindent
The resulting interfacial tension force can be rewritten as the sum of each component force

\begin{equation}\label{eqB4x}
\mathbf{f}_{st}^\ast =\sum_{i=1}^{N_c} \Sigma_i^\ast \kappa_i^\ast \mathbf{n}_i \delta_i^\ast \, .
\end{equation}

\noindent
In accordance with the aforementioned works, we also use $\delta_i^\ast = |\nabla \alpha_i |$, $\mathbf{n}_i = \nabla\alpha_i/ (| \nabla\alpha_i | + sf)$ and $\kappa_i^\ast=\nabla \cdot \mathbf{n}_i$, and the continuum surface force can be written based on the component volume fraction $a_i$ as:

\begin{equation}\label{eqB4}
\mathbf{f}_{st}^\ast =\sum_{i=1}^{N_c} \Sigma_i^\ast \kappa_i^\ast \nabla\alpha_i \, .
\end{equation}

\noindent
It is noted that according to this formulation the unit normal $\mathbf{n}_i$ and the curvature $\kappa_i^\ast$ have to be calculated separately for each phase $i$.

\subsubsection{VOF smoothing} \label{sec2.2.2}
In the VOF method, the fluid interface is implicitly represented by the volume fraction, with its value changing abruptly in a small but finite region. This creates errors in the evaluation of the normal vectors and the curvature of interfaces, which are used to calculate the interfacial body forces. In flows dominated by interfacial tension effects, these errors usually give rise to non-physical parasitic currents adjacent to the interface due to local variations in the continuum surface force \cite{Lafaurie1994, Scardovelli1999, Harvie2006}. These numerical artifacts may be efficiently suppressed by evaluating the curvature of interfaces $\kappa_i^\ast$ from a smoothed VOF distribution $\tilde{a_i}$. Here, the smoothed function is calculated from $a_i$ by applying a filter over a finite region around the interface \cite{Scardovelli1999, Gerlach2006}. Thus, the curvature $\kappa_i^\ast$ is computed as

\begin{equation}\label{eqC1}
\kappa_i^\ast = \nabla \cdot \left( \frac{\nabla \tilde{\alpha_i}}{\vert\nabla \tilde{\alpha_i} \vert \, + \, sf} \right) \, .
\end{equation}

\noindent
We applied as a smoother a Laplacian filter proposed by Lafaurie et al. \cite{Lafaurie1994} that transforms the original VOF function $a_i$ into a smoothed function $\tilde{a_i}$

\begin{equation}\label{eqC2}
\tilde{\alpha}_{i,\mathrm{p}} = \frac{\sum_{\mathrm{f}=1}^{nf} (\alpha_{i})_\mathrm{f} |\mathbf{S}_\mathrm{f}^\ast|}{\sum_{\mathrm{f}=1}^{nf} |\mathbf{S}_\mathrm{f}^\ast|} \, ,
\end{equation}

\noindent
where the subscript '$\mathrm{p}$' denotes the cell index and '$\mathrm{f}$' denotes the face index, $nf$ is the number of faces of a finite control volume, and $\mathbf{S}_\mathrm{f}^\ast$ is the cell surface area. The interpolated value $(a_{i})_\mathrm{f}$ at the face centre is calculated using linear interpolation. The application of this filter can be repeated $n_{lfi}$ times in order to obtain a smoothed field. However, it is pointed out that the smoothing operation should be applied up to the level that is necessary to effectively suppress the parasitic currents, as it generally tends to level down regions of the interface having high curvature. It is noted that the smoothed VOF distribution $\tilde{a_i}$ is only used in the evaluation of the curvature (see Eq. \ref{eqC1}) and the regular VOF function $a_i$ is used in all other equations.

\subsubsection{Interface sharpening} \label{sec2.2.3}

In the new \texttt{multiFluidInterFoam} solver, the fluid interfaces are also sharpened by introducing artificial compression terms into Eq. (\ref{eqA4}) in a similar manner as in the original \texttt{multiphaseInterFoam} solver. For consistency reasons, though, the artificial compression velocity $\mathbf{u_r}_{,ik}^\ast$ is now calculated as 

\begin{equation}
(\mathbf{u_r}_{,ik}^\ast)_\mathrm{f} = (\mathbf{n}_{i})_\mathrm{f} \min \left[ C_{a_i} \frac{|\phi_{\mathrm{f}}^\ast|}{|\mathbf{S}_\mathrm{f}^\ast|}, \max \left( \frac{|\phi_{\mathrm{f}}^\ast|}{|\mathbf{S}_\mathrm{f}^\ast|}\right) \right], \,\,\,\,\,\,\,\,\,\,\,\,  i=1, 2,..., N_c \,\,\, \mathrm{and} \,\,\, k \ne i \, ,
\label{VOFComp2n}
\end{equation}

\noindent
where the normal vector $(\mathbf{n}_{i})_\mathrm{f}$ is used instead of $(\mathbf{n}_{ik})_\mathrm{f}$, and it is computed as 

\begin{equation} \label{VOFComp4}
(\mathbf{n}_{i})_\mathrm{f} = \frac{(\nabla\alpha_{i})_\mathrm{f}} {\mid (\nabla \alpha_{i})_\mathrm{f} \mid \, + \, sf } \, ,
\end{equation}

\noindent
similarly to the evaluation of the normal vector in the calculation of the continuum surface force Eq. (\ref{eqB4}) in the new solver.

\subsubsection{\texttt{SIMPLEC} velocity-pressure coupling} \label{sec2.2.4}

As it will be discussed below in detail, the implemented version of the \texttt{PIMPLE} algorithm for the velocity-pressure coupling in the \texttt{multiphaseInterFoam} solver was not sufficiently stable for transient viscoplastic simulations in the floating lens, drop levitation and rising bubble problems. 
In order to avoid numerical instabilities or divergence, we found that it was necessary to (a) under-relax pressure and velocity fields, (b) use a sufficiently large number of inner iterations performing the velocity-pressure coupling, or (c) use a very small time-step. The first option is not efficient for transient simulations, while the second and third ones lead to a burden on the computational cost.

The \texttt{SIMPLEC} (Semi-Implicit Method for Pressure-Linked Equations-Consistent), algorithm \cite{Doormaal1984} in opposition to the \texttt{SIMPLE}, does not require under-relaxation of pressure and velocity, except for simulations with non-orthogonal grids. Additionally, its computational cost per iteration is lower than in the \texttt{PISO} algorithm, because the pressure equation is only solved once per cycle. For these reasons, the \texttt{SIMPLEC} algorithm was adopted in the present work. Note that the \texttt{SIMPLEC} algorithm is already available in the \texttt{OpenFOAM} as a possible steady-state solver for (in)compressible, inelastic fluids. The \texttt{SIMPLEC} method is implemented in the new multiphase solver in a similar manner as in the solvers provided by the \texttt{RheoTool} \cite{rheoTool2016} toolbox and, thus, the details of its implementation are not repeated here.

\subsection{Overview of the \texttt{rheoMultiFluidInterFoam} solver}\label{sec2.3}

The \texttt{rheoMultiFluidInterFoam} solver is an extension of the \texttt{multiFluidInterFoam} solver, in which the library containing the non-Newtonian fluid models of \texttt{RheoTool} \cite{rheoTool2016} has been fully incorporated enabling the modelling of a wide variety of complex fluids, e.g. exhibiting viscoelastic, viscoplastic or elasto-viscoplastic effects. It is recalled that for the purposes of the present work we mainly focus on the simulation of Newtonian, viscoplastic and viscoelastic fluids.

\subsubsection{Viscoplastic material - Herschel--Bulkley model}\label{sec2.3.1}
A viscoplastic material deforms as non-Newtonian viscous fluid when the local stress exceeds some critical value, whereas it behaves as a solid at low values of the stress. The discontinuous Herschel--Bulkley constitutive equation is commonly used for modelling a viscoplastic fluid without accounting for the elasticity of the material and is given by

\begin{equation}\label{eqD1}
\mathbf{\underline{\tau}}^\ast = \left(  \tau_0^\ast/\dot{\gamma}^\ast+ k^\ast (\dot{\gamma}^\ast)^{n-1}  \right) \mathbf{\dot{\underline{\gamma}}^\ast}, \,\,\,\,\,\,  |\mathbf{\underline{\tau}}^\ast| \ge \tau_0^\ast \, ,
\end{equation}

\begin{equation}\label{eqD2}
\dot{\gamma}^\ast=0,  \,\,\,\,\,\,  |\mathbf{\underline{\tau}}^\ast| < \tau_0^\ast \, ,
\end{equation}

\noindent
where $|\mathbf{\underline{\tau}}^\ast|=\sqrt{(\mathbf{\underline{\tau}}^\ast:\mathbf{\underline{\tau}}^\ast)/2}$; the exponent $n$ is the power law index and $k^\ast$ is a consistency factor. In order to overcome the discontinuity introduced in the constitutive equation by the separation of yielded and unyielded regions, commonly  viscosity regularization models and methods based on variational principles have been used \cite{Balmforth2014, Mitsoulis2017}.

The most famous variational approach for the solution of viscoplastic problems is the Augmented Lagrangian method \cite{Dimakopoulos2013, Dimakopoulos2018, Glowinski1984, Vinay2006}, allowing one to cope with the exact discontinuous law. A popular alternative to the augmented Lagrangian approach are regularization models, where the discontinuous viscosity function is smoothed in order to asymptotically match the original model. Arguably the most commonly used approximation among them is the Papanastasiou's regularization \cite{Papanastasiou1987, Tsamopoulos2008, Tripathi2015}. The main advantage of this approach is that it is applied easily in fluid flow solvers. Despite the simplicity of the Papanastasiou regularization model and the fact that it is generally found to be superior than bi-viscosity models \cite{Frigaard2005}, it is not currently available in \texttt{OpenFOAM}. It is provided, though, by the \texttt{RheoTool}, which is incorporated in the new Non-Newtonian multiphase solver, presented herein. The Papanastasiou regularization model can be written as

\begin{equation}\label{eqD1pap}
\nu_i^\ast \left(\dot{\gamma}^\ast\right) = \min \left(  \nu_{0,i}^\ast, \frac{\tau_{0,i}^\ast} {\dot{\gamma}^\ast} \left[ 1-\exp(-m_i^\ast \dot{\gamma}^\ast)\right] + k_i^\ast (\dot{\gamma}^\ast)^{n_i-1} \right) \, ,
\end{equation}

\noindent
where $m_i^\ast$ is an additional parameter related to the exponential term. It is noted that the original Papanastasiou regularization does not include the artificial upper-bounding by $\nu_{0,i}^\ast$. However, this bounding is needed in order to avoid an infinite viscosity for $\dot{\gamma}^\ast\longrightarrow 0$ (e.g. startup of flow) and $n_i < 1$. The original Papanastasiou regularization is recovered for $\nu_{i,0}^\ast \longrightarrow \infty$. In practice, $\nu_{0,i}^\ast$ should be low enough to avoid an infinite viscosity in quiescent conditions and high enough to allow Papanastasiou regularization to take control in the remaining situations. It is noted that the model Eqs. (\ref{eqA3a}) and (\ref{eqA3}) used in both the original \texttt{multiphaseInterFoam} solver and the new \texttt{multiFluidInterFoam} solver are also available in the \texttt{rheoMultiFluidInterFoam} solver through the \texttt{RheoTool} toolbox.

We note that the \texttt{rheoMultiFluidInterFoam} solver adopts a different approach for the calculation of the deviatoric stress tensor $\mathbf{\underline{\tau}}^\ast$ in Eq. (\ref{eqA1}) with respect to the \texttt{multiFluidInterFoam} and \texttt{multiphaseInterFoam} solvers. In Eq. (\ref{eqA3b}) used in the latter solvers, a bulk density and a bulk kinematic viscosity are required that are calculated as weighted averages of the density and viscosity of the involved generalized Newtonian fluids, respectively. Currently, the \texttt{rheoMultiFluidInterFoam} solver solves a constitutive equation for each phase and the extra-stress tensor contributing to the momentum equation is the weighted average of the extra-stress tensor for each phase $\mathbf{\underline{\tau}}_i^\ast$ as follows

\begin{equation}\label{eqA3bx}
\mathbf{\underline{\tau}}^\ast = \sum_i a_i \mathbf{\underline{\tau}}_i^\ast \, .
\end{equation}

\subsubsection{Viscoelastic material - Phan-Thien Tanner model}\label{sec2.3.2}

Several constitutive equations are available in \texttt{RheoTool} to model the viscoelastic properties of complex fluids. In this work, the linear Phan-Thien Tanner (PTT) viscoelastic model was chosen in order to assess the performance of the newly developed solver to predict complex fluid flows. This model reproduces both the elastic behavior and the dependendence of the shear and extensional viscosity on local rate of deformation of several fluids. To simulate viscoelastic fluid flows, in particular, the total extra-stress tensor is usually split into a solvent contribution ($\mathbf{\underline{\tau}}_s^\ast$) and a polymeric contribution ($\mathbf{\underline{\tau}}_p^\ast$), so that Eq. (\ref{eqA3bx}) can be re-written as 
\begin{equation}\label{eqA3bx1a}
\mathbf{\underline{\tau}}^\ast = \sum_i a_i (\mathbf{\underline{\tau}}_{p,i}^\ast + \mathbf{\underline{\tau}}_{s,i}^\ast)\, .
\end{equation}

\noindent
A constitutive equation is required for each tensor contribution, which for the linear PTT model can be written as

\begin{equation}\label{eqA3bx1}
\mathbf{\underline{\tau}}_{s,i}^\ast = \mu_{s,i}^\ast \left(  \dot{\gamma}^\ast \right) \mathbf{\dot{\underline{\gamma}}^\ast} \, ,
\end{equation}

\begin{equation}\label{eqA3bx2}
\left[ 1+\frac{\epsilon_i \lambda_i^\ast}{\mu_{p,i}^\ast}\mathrm{tr}(\mathbf{\underline{\tau}}_{p,i}^\ast) \right]  \mathbf{\underline{\tau}}_{p,i}^\ast + \lambda_i^\ast \left( \dot{\gamma}^\ast \right) 
\accentset{\circ}{\mathbf{\underline{\tau}}}_{p,i}^\ast  = \mu_{p,i}^\ast \left( \dot{\gamma}^\ast \right) \mathbf{\dot{\underline{\gamma}}^\ast} \, ,
\end{equation}

\noindent
where $\mu_{s,i}^\ast$ is the solvent viscosity, $\mu_{p,i}^\ast$ is the polymeric viscosity coefficient, $\lambda_i^\ast$ is the relaxation time, $\epsilon_i$ is the extensibility parameter controlling the degree of shear and extensional thinning, and 
\begin{equation}
\accentset{\circ}{\mathbf{\underline{\tau}}}_{p,i}^\ast = \accentset{\nabla}{\mathbf{\underline{\tau}}}_{p,i}^\ast + \frac{\zeta}{2} (\mathbf{\underline{\tau}}_{p,i}^\ast \cdot \mathbf{\dot{\underline{\gamma}}^\ast} + \mathbf{\dot{\underline{\gamma}}^\ast} \cdot \mathbf{\underline{\tau}}_{p,i}^\ast)
\end{equation}
is the Gordon--Schowalter derivative, which is controlled by the parameter $\zeta$ in order to take non-affine deformation into account. 
The term 
\begin{equation}
\accentset{\nabla} {\mathbf{\underline{\tau}}}_{p,i}^\ast = \frac{\partial \mathbf{\underline{\tau}}_{p,i}^\ast}{\partial t^\ast} + \mathbf{u}^\ast \cdot \nabla \mathbf{\underline{\tau}}_{p,i}^\ast - \mathbf{\underline{\tau}}_{p,i}^\ast \cdot \nabla \mathbf{u}^\ast - \nabla \mathbf{u}^{\ast \mathrm{T}} \cdot \mathbf{\underline{\tau}}_{p,i}^\ast
\end{equation}
represents the upper-convected time derivative, which renders the model frame-invariant. The log-conformation approach is utilized in order to solve numerically Eq. (\ref{eqA3bx2}). For more details, please refer to Reference \cite{rheoTool2016}.

\subsection{Numerical methods and details}\label{sec2.4}

In the developed multiphase solvers, the time discretization is performed by using a second-order Crank--Nicolson scheme, except for the phase fraction equations in which a first-order Euler method is used. The space discretization is based on a second-order linear scheme. The advection of velocity is discretized by utilizing the \textit{limitedLinearV 1} scheme, while the \textit{vanLeer01} and \textit{interfaceCompression} schemes are used for the phase fractions. The resulting linear system of equations is solved by using the Preconditioned bi-conjugate gradient solver with Diagonal incomplete-LU (PBiCG/DILU) for velocity, the Pre-conditioned Conjugate Gradient solver combined with Diagonal Incomplete Cholesky (PCG/DIC) for pressure and pressure correction, and a smoothSolver/Gauss-Seidel solver for the phase fractions. For all variables, the tolerance was set equal to $10^{-12}$, whereas no under-relaxation was used. The time step was determined using a global and local Courant numbers of $0.5$, while the maximum allowed time step was determined as the smallest obtained from the equations \cite{Personnettaz2018, Deshpande2012}

\begin{equation} \label{dt1}
\Delta t_{c,i}^\ast = C_{c,i} \sqrt{ \rho_i^\ast \Delta x^{\ast \, 3} / |\Sigma_i^\ast|} \,\,\,  i=1,2,3,
\end{equation}

\begin{equation} \label{dt2}
\Delta t_{v,i}^\ast = C_{v,i} \mu_i^\ast \Delta x^\ast / |\Sigma_i^\ast| \,\,\,  i=1,2,3,
\end{equation}

\noindent
where $C_{c,i} = C_{v,i} = 0.5$. The above settings were used in this work unless otherwise stated.

\subsection{Overview of the new solvers}\label{sec2.5}

The computational steps of the new multiphase solvers are as follows:

\begin{enumerate}
\item Initialize all the fields $\lbrace p^\ast, \mathbf{u}^\ast, a_i \rbrace_0$ and time $(t^\ast = 0)$

\item Enter the time loop $(t^\ast=\Delta t^\ast)$

	\begin{enumerate}
	\item Enter the inner iterations loop $(it=0)$

		\begin{enumerate}
		\item Calculate the Courant numbers and adjust the time step based on the imposed restraints if necessary 

		\item Solve the $a_i^{(it)}$ equations by using the volumetric fluxes of the previous time level

		\item For a generalized Newtonian fluid, use the appropriate constitutive equations together with the new values of $a_i^{(it)}$ in order to update the estimates for the cell center viscosity $\mu^{\ast \, (it)}$ and density $\rho^{\ast \, (it)}$ and the face viscosity $\mu_{f}^{\ast \, (it)}$ and density $\rho_{f}^{\ast \, (it)}$ of the multiphase mixture. For a viscoelastic material, solve a differential constitutive equation for each phase and compute the extra-stress tensor contributing to the momentum equation as the weighted average of the extra-stress tensor for each phase by using the indicator VOF function $a_i^{(it)}$; the latter approach can be used for every rheological model described by a differential constitutive equation, e.g. for elasto-viscoplastic materials

\item Based on the above values, solve the momentum equation to obtain a velocity prediction $\mathbf{u}^{\ast \, (it)}$

\item Continue with the pressure-correction algorithm (SIMPLEC) and solve the pressure equation to obtain a continuity-compliant pressure field, $p^{\ast \, (it)}$

\item Correct both the face $\mathbf{\hat{u}}_f^{\ast \, (it)}$ and cell-centered $\mathbf{\hat{u}}^{\ast \, (it)}$ velocities using the correct pressure field

\item Increment the inner iteration index $(it=it+1)$ and return to step $i$ until the pre-defined number of inner iterations is reached

\item Set $\lbrace p^\ast, \mathbf{u}^\ast, a_i \rbrace_t = \lbrace p^{\ast \, (it)}, \mathbf{u}^{\ast \, (it)}, a_i^{(it)} \rbrace$ 
\end{enumerate}

\item If the final time has not been reached, advance to the next time level $t=t+\Delta t$ and return to step $2(a)$
\end{enumerate}

\item Stop the simulation and exit
\end{enumerate}

\noindent
The only difference between the two new solvers is in step $2(a)iii$, in which the \texttt{multiFluidInterFoam} solver uses constitutive relations like Eqs. (\ref{eqA3a}) and (\ref{eqA3}) already available in \texttt{OpenFOAM}, whereas the \texttt{rheoMultiFluidInterFoam} solver employs constitutive relations given by the \texttt{RheoTool}. In the above numerical algorithm, there is a main loop advancing in time, step 2, and an inner loop, step 2(a). The purpose of the inner iterations is to reduce the explicitness of the numerical method, which is crucial to enhance both the stability and the time accuracy of the solver. The lastly computed value of a variable is used in the inner loop, either from previous iteration of the inner loop or from the previous time step.

The main changes introduced relative to the original \texttt{multiphaseInterFoam} solver can be summarized as follows: (a) the interfacial tension coefficient decomposition method is employed to deal with tension pairings between different faces using a compositional approach, and the resulting interfacial force is reformulated properly as the sum of each component body force, (b) VOF smoothing is implemented to estimate accurately the curvature of interfaces, (c) the velocity-pressure coupling is utilized based on the \texttt{SIMPLEC} algorithm, and (d) the \texttt{RheoTool} toolbox has been implemented in the new solver to make use of several constitutive models for complex fluids. 

\section{Results}\label{sec3}

We begin our study in Section \ref{sec3.1} with the investigation of the behavior of our new solver \texttt{multiFluidInterFoam}. To this end, the results obtained by the new solver are compared against those yielded by the original \texttt{multiphaseInterFoam} solver for three benchmark flow problems: (a) the three-phase flow in a dam break with an obstacle, (b) the spreading of a droplet between two stratified fluids, and (c) the levitation of a droplet between two stratified fluids. Case (a) is a slightly modified case with respect to the standard \texttt{damBreak4Phase} tutorial (i.e. a flow case with four fluid phases) available in \texttt{OpenFOAM}. 

Next, we proceed in Section \ref{sec3.2} with the validation of the \texttt{rheoMultiFluidInterFoam} solver. Given that there are no available benchmarks in the literature for Non-Newtonian three-phase flows and in order to validate the code and the proper implementation of \texttt{RheoTool}, we consider the limiting case of the two-phase flow of a Newtonian bubble rising in (a) a viscoplastic fluid described by the Herschel--Bulkley model and (b) a viscoelastic fluid described by the Phan-Thien Tanner model. The former case is useful since it is possible to perform a comparison between the two new multiphase solvers when using the bi-viscosity version of the Herschel--Bulkley model, while it is also possible to assess the numerical implementation of the Herschel--Bulkley model based on the \texttt{RheoTool}. On the other hand, the latter case provides a representative test for the implementation of differential constitutive equations provided by the \texttt{RheoTool}.

\subsection{Consistency and verification tests for the \texttt{multiFluidInterFoam} solver} \label{sec3.1}


\subsubsection{Dam break with obstacle} \label{sec3.1.1}
The newly developed solver is capable of predicting the multiphase flow of three or more immiscible fluids by capturing moving interfaces on an Eulerian mesh. Merging or breakup of the interfaces is also handled in a natural way. A modified and refined version of the standard dam break tutorial of \texttt{OpenFOAM} is used here as an initial investigation of the behavior of our new solver; for our comparison we consider the case of a three-phase system.

The considered computational domain is a square box having width and height of $0.584\,\mathrm{m}$.
A small rectangular obstacle of size of $0.284\, \mathrm{m} \times 0.484 \, \mathrm{m}$ is located on the bottom wall at a distance of $0.484 \, \mathrm{m}$ from the left wall. Phases $1$ (water, blue color) and $2$ (oil, green color) are initially placed in square columns at the left side of the domain, while the rest of the box is filled with phase $3$ (air, red color). The dynamic viscosity and density of phase $1$ are $10^{-3} \,\mbox{Pa s}$ and $1000 \,\mathrm{kg} \, \mathrm{m}^{-3}$, respectively, while those of phases $2$ and $3$ are $1 \times 10^{-6}\,\mbox{Pa s}$, $500 \,\mathrm{kg} \, \mathrm{m}^{-3}$ and $1.48 \times 10^{-5} \,\mbox{Pa s}$, $1 \,\mathrm{kg} \, \mathrm{m}^{-3}$, respectively. The surface tension coefficients for all interfaces are set to $0.07\,\mathrm{N}\,\mathrm{m}^{-1}$. The magnitude of gravity acceleration is $g^\ast=9.81 \, \mathrm{m}\, \mathrm{s}^{-2}$. 

No-slip boundary condition is applied to all boundaries except for the top surface, which is assumed to be an open boundary. Zero gradient is applied at the walls for each volume fraction, while at the open boundary all fractions are assigned to zero expect for that of phase $3$ which is set to unity. The resolution of the computational mesh is $92 \times 92$ control volumes. All the simulations use the same computational mesh and discretizations schemes summarized in Section \ref{sec2.4}. The time step is determined using a global and local Courant numbers of $0.5$, while the maximum allowed time step is set equal to $\Delta t^\ast=10^{-4}\,\mathrm{s}$. The interface compression parameter $C_{a_i}$ is set to unity. No smoothing in the calculation of curvature is applied in these simulations to facilitate the comparison with the original multiphase solver.

\begin{figure}[h!]
\centerline{\includegraphics[width=0.95\linewidth] {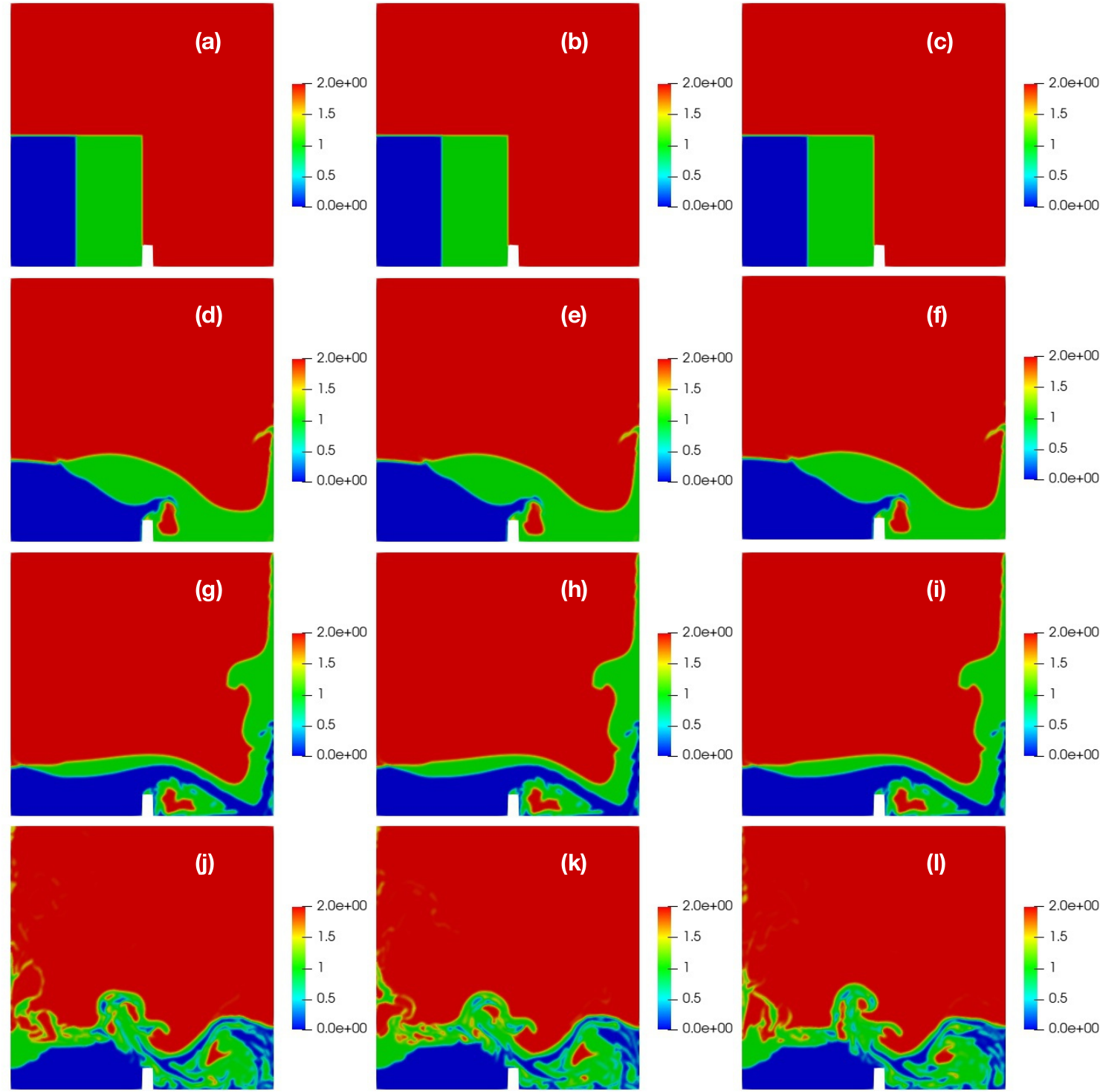}}
\caption{Distribution of phase $1$ (water, blue color), phase $2$ (oil, green color) and phase $3$ (air, red color) for the problem of dam break with an obstacle at times $t^\ast=0\,\mathrm{s}$ [(a), (b), (c)], $t^\ast=0.25\,\mathrm{s}$ [(d), (e), (f)], $t^\ast=0.5\,\mathrm{s}$ [(g), (h), (i)], and $t^\ast=1\,\mathrm{s}$ [(j), (k), (l)]. The results are obtained by using the new \texttt{multiFluidInterFoam} solver with the \texttt{SIMPLEC} method (left column) or the \texttt{PIMPLE} method (middle column) for velocity-pressure coupling, and the original \texttt{multiphaseInterFoam} solver (right column).}
\label{figDam1}
\end{figure}

The initial condition is given in Figs. \ref{figDam1}(a),(b),(c). The simulations were performed based on the new \texttt{multiFluidInterFoam} solver by using either the \texttt{SIMPLEC} (see Figs. \ref{figDam1}(a),(d),(g),(j)) or the \texttt{PIMPLE} (see Figs. \ref{figDam1}(b),(e),(h),(k)) pressure-velocity coupling algorithms, and the original \texttt{multiphaseInterFoam} solver (see Figs. \ref{figDam1}(c),(f),(i),(l)) for a time period of $1 \, \mathrm{s}$. It can be seen that all solvers give almost the same prediction for the distribution of the phases, as shown in Figs. \ref{figDam1}(d)-(i). As time proceeds, the three phases change their shape and position because of density differences. At time $t^\ast=0.25\,\mathrm{s}$, phase $1$ has just stumbled over the obstacle on the floor, an air bubble is trapped adjacent to the obstacle, whereas phase $2$ has hit the right wall and moves upward. At time $t^\ast=0.5\,\mathrm{s}$, the air bubble is located within a larger drop of phase $2$, while phase $1$ has reached the right wall. Overall, we observe that the qualitative characteristics of the distribution of phases are similar in all three simulations, with small differences arising only at the very late stages of the simulation, at $t^\ast=1\,\mathrm{s}$. This kind of multiphase flow depends mostly on the density differences between the phases, which generate sufficient momentum so that any surface tension effects are rather small. 
By using the height of the fluid columns as a length scale $H^\ast=0.292\,\textrm{m}$, a proper characteristic velocity scale can be determined as $(\rho_1^\ast-\rho_2^\ast)/\rho_2^\ast(g^\ast H^\ast)^{1/2}$ and, thus, the Weber number can be estimated as $We=\rho_1^\ast U^{\ast 2} H^\ast/\sigma_{12}^\ast \approx 1.2 \times 10^{4}$, where $\rho_1^\ast$, $\rho_1^\ast$ are the density of phases $1$ and $2$, respectively, and $\sigma_{12}^\ast$ is the interface tension coefficient of the interface between fluids $1$ and $2$. The Weber number measures the relative magnitude of the inertia and surface tension forces, and the presently high $We$ value indeed suggests that any surface tension effects are negligible and the phenomenon is driven mostly by the inertia forces. Clearly in this example, the contact point, that is observed at the early times of the simulation and moves toward the left wall, do not control the dynamics of the multiphase flow, which exhibits interfaces mainly between two phases.

\subsubsection{Spreading of a droplet between two stratified Newtonian fluids}\label{sec3.3}
The spreading of a liquid lens is a very classic benchmark problem and it has been widely used previously to assess the numerical methods for three-phase flows \cite{Smith2002, Xie2020, Garcke1999, Kim2004, Kim2007, Leclaire2013}. Here, we examine this benchmark problem in detail and consider the cases of partial and full spreading of the droplet in order to investigate whether the newly developed numerical solver is capable of accurately predicting the dynamics of spreading and also producing the correct equilibrium contact angles with triple junction points.

\subsubsection*{\it Problem description} \label{sec3.3.1}

\begin{figure}[b!]
\centerline{\includegraphics[width=0.5\linewidth]{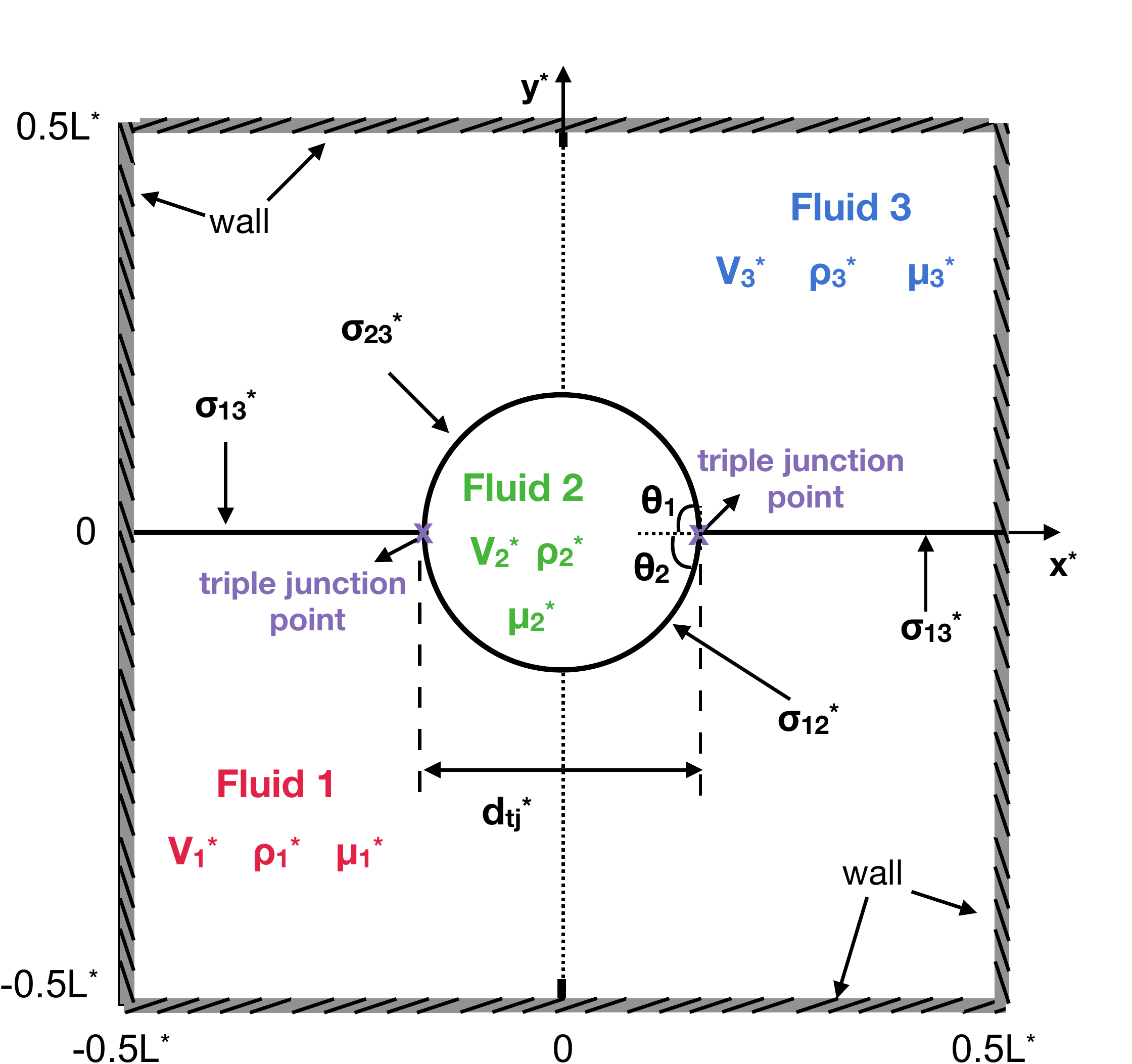}}
\caption{Geometry and initial setup of the spreading of a droplet between two stratified fluids problem.}
\label{figLensGeo}
\end{figure}

The geometry and relevant variables for the spreading of a droplet between two stratified fluids are shown in Fig. \ref{figLensGeo}. We consider the case of a viscous fluid $2$ of density $\rho^\ast_2$ and viscosity $\mu^\ast_2$ spreading between two fluid layers of density $\rho^\ast_1$, $\rho^\ast_3$ and viscosity $\mu^\ast_1$, $\mu^\ast_3$. The bottom and upper layers are bounded from below and above as well as from left and right by impermeable, rigid, and solid substrates. The surface tension of the interface between fluids $i$ and $j$ is denoted with $\sigma^\ast_{ij}$ ($i=1,2,3$ and $j=1,2,3$). All fluids are assumed to be incompressible, Newtonian and immiscible.

Initially, a circular droplet (fluid $2$) with radius $R_d^\ast$ and area $A^\ast$ is placed between the interface of the lower fluid $1$ and upper fluid $3$ with zero velocity. The center of the droplet is located at the point $(X_c^\ast, \, Y_c^\ast)$. No-slip boundary condition is applied along the walls, 
whereas a symmetry condition is applied at the left boundary; therefore, the computational domain consists of half of the square box shown in Fig.\ref{figLensGeo}. The characteristic scales of length, velocity, time and pressure are denoted as $L^\ast$, $U^\ast$, $U^\ast / L^\ast$ and $\rho^\ast U^{\ast 2}$, respectively. The total height of the fluid layers is chosen as a proper length scale $L^\ast$, while the characteristic velocity is defined as $U^\ast = \mu^\ast_{1}/(\rho^\ast_1\,L^\ast)$. Here, the gravity effects are not taken into account. The problem is characterized by the non-dimensional density, viscosity, and surface tension coefficients given as $\rho_i = \rho_i^\ast/\rho^\ast_1$, $\mu_i = \mu^\ast_i/\mu^\ast_1$, and $\sigma_{ij} = \sigma_{ij}^\ast/\sigma^\ast_{23}$, $\Sigma_i = \Sigma_i^\ast/\Sigma^\ast_2$ ($i,j=1,2,3$), respectively. The Reynolds number is defined as $Re = \rho^\ast_1 U^\ast L^\ast/\mu^\ast_1$, and the component Reynolds number for each  $i$ phase $Re_i=\rho_i/\mu_i Re$. The Weber number is given as $We=\rho^\ast_1 L^\ast U^{\ast 2}/\sigma^\ast_{23}$ and the component Weber number as $We_{ij}= \rho_i/\sigma_{ij} We$, while the phasic Weber number is given as $We_p=\rho^\ast_1 L^\ast U^{\ast 2}/\Sigma^\ast_2$ and the component phasic Weber as number $We_{p,i}= \rho_i/\Sigma_i We_p$.

Two different spreading phenomena can be observed depending on the values of the interfacial tensions, i.e., partial spreading and full spreading. This can be categorized based on the spreading parameter of the phase $i$ at the interface between phases $j$ and $k$ \cite{Karapetsas2011, Craster2006, Boyer2010}

\begin{equation}\label{Si}
S_i^\ast=\sigma_{jk}^\ast-\sigma_{ij}^\ast-\sigma_{ik}^\ast.
\end{equation}

\noindent
If $S_i^\ast$ is negative $(S_i^\ast<0)$, the spreading is said to be partial. In this case, each of the three interfacial tension coefficients is less than the sum of the other two coefficients and, thus, stable triple points can be observed in regions where the three fluids meet. On the other hand, if $S_i^\ast$ is positive $(S_i^\ast>0)$, the spreading is said to be full or complete. In this case, one interfacial tension is so large that the sum of the other two cannot balance it. The triple points become unstable and one fluid will spread in a such a way that the triple points disappear. Then, the system tends to an equilibrium state in which the interfaces are finally either plane or spherical, depending on the combination of values of the interfacial tension coefficients.


\subsubsection*{\it Partial spreading} \label{sec3.3.2}
 
We first consider a case resulting in partial spreading, for which the three interfacial tensions yield a Neumann's triangle. In partial spreading, the droplet can eventually reach a steady lens shape having geometrical characteristics that can be analytically calculated \cite{Rowlinson2013, Liang2016, Yu2019}. For the cases presented here, we consider a droplet with initial radius of $R_d^\ast=150 \, \mu \mbox{m}$ and it is placed at $(X_c^\ast,Y_c^\ast)=(0,0)$, while the characteristic length scale is $L^\ast=1\, \textrm{mm}$. %
All fluids have the same density and dynamic viscosity and therefore $\rho_i=\mu_i=1$ ($\rho_1^\ast=1000 \, \textrm{Kg\,m}^{-3}$, $\mu_1^\ast=1 \textrm{Pa\,s}$) for $i=1,2,3$. The Reynolds number is $Re=Re_1=Re_2=Re_3=1$. The phasic Weber number corresponding to the phase $2$ is kept constant to $We_{p,2} = 60$ while we consider three different cases of increasing interfacial tension with $We_{p,1}=We_{p,3}=108\,(\sigma_{23}^\ast=0.026\,\textrm{N\,m}^{-1})$, $60\,(\sigma_{23}^\ast=0.033\,\textrm{N\,m}^{-1})$, $36\,(\sigma_{23}^\ast=0.044\,\textrm{N\,m}^{-1})$. We also vary the parameters $C_{a_i}$ and $n_{lfi}$ to examine the effect of interface sharpening and smoothing operation in the calculation of curvature, respectively. All the simulations use the same computational mesh, discretizations schemes and numerical parameters. The computational mesh consists of $100 \times 200$ control volumes, unless otherwise stated.

\begin{figure}[b!]
\centerline{\includegraphics[width=1\linewidth] {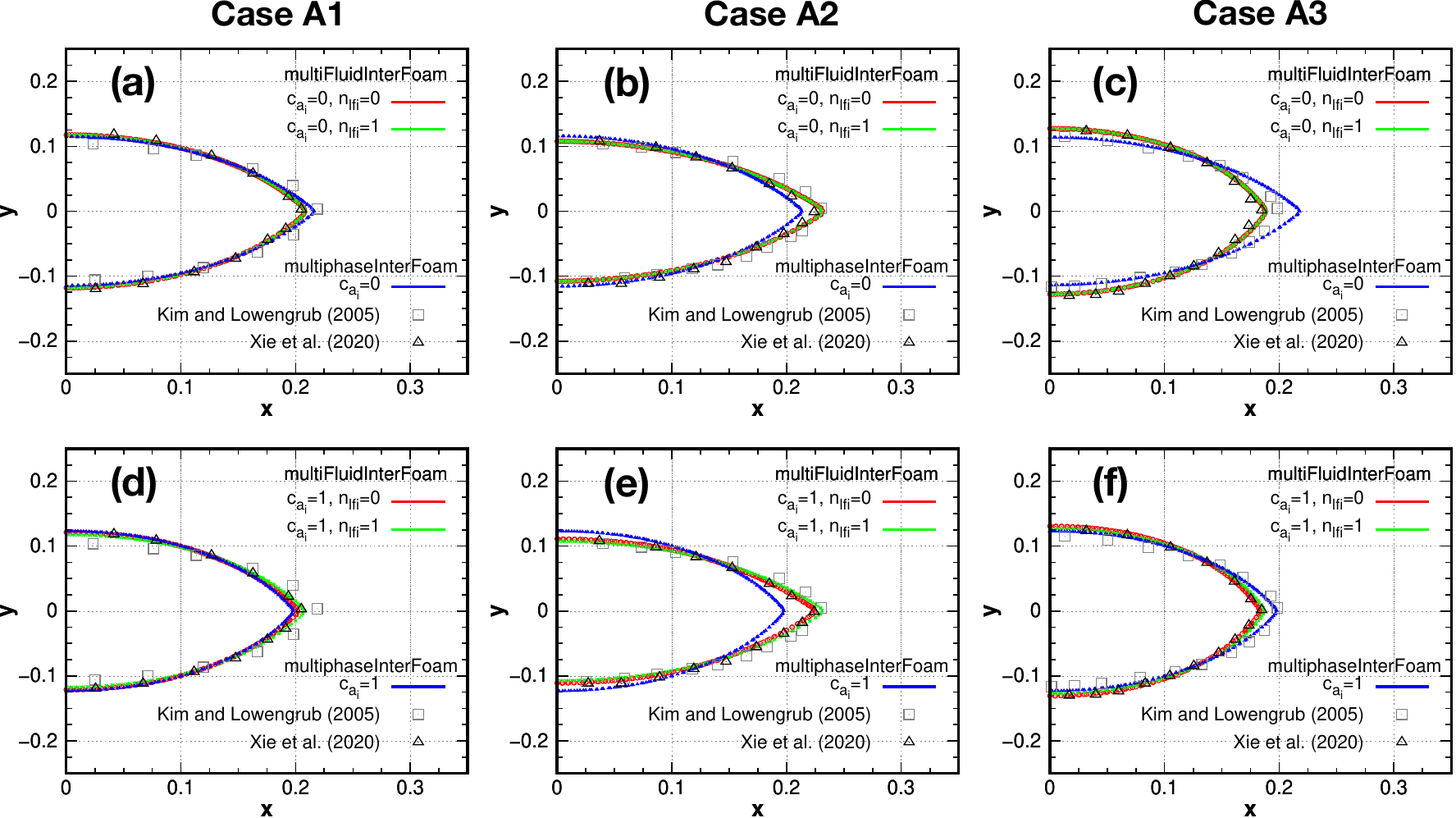}}
\caption{Equilibrium shape of the droplet (corresponding to $a_2=0.5$ isoline) predicted by the new \texttt{multiFluidInterFoam} solver for Case A$1$ with $We_{p,1}=We_{p,2}=We_{p,3}=60$ (left column), Case A$2$ with $We_{p,2}=60$ and $We_{p,1}=We_{p,3}=36$ (middle column), and Case A$3$ with $We_{p,2}=60$ and $We_{p,1}=We_{p,3}=108$ (right column). Simulations are performed with various combinations of the interface compression parameter and the filtering parameter $C_{a_i}=0$, $n_{lfi}=0,1$ [(a),(b),(c)] and $C_{a_i}=1$, $n_{lfi}=0,1$ [(d),(e),(f)]. The results are compared with the numerical results of Xie et al. \cite{Xie2020} and Kim and Lowengrub \cite{Kim2005} and those obtained by the regular \texttt{multiphaseInterFoam} solver. The density and viscosity ratios are $\rho_i=1$ and $\mu_i=1$, respectively.}
\label{figA1}
\end{figure}

In Fig.\ref{figA1}, we show the equilibrium droplet shapes for three different cases (i.e. A1, A2 and A3) as predicted by our new \texttt{multiFluidInterFoam} solver for various combinations of the interface compression parameter $C_{a_i}$ and the Laplacian filter parameter $n_{lfi}$. Our solutions are compared with previous numerical results of Xie et al. \cite{Xie2020} and Kim and Lowengrub \cite{Kim2005} and those obtained by using the original \texttt{multiphaseInterFoam} solver. Case A$1$ corresponds to $We_{p,1}=We_{p,2}=We_{p,3}=60$ and is shown in Figs. \ref{figA1}(a),(d); Similarly, in Figs. \ref{figA1}(b),(e) we depict the predictions for Case A$2$ with $We_{p,2}=60$ and $We_{p,1}=We_{p,3}=36$, 
and we show the results for Case A$3$ with $We_{p,2}=60$ and $We_{p,1}=We_{p,3}=108$ in Figs. \ref{figA1}(c),(f). In the panels (a)-(c), the numerical parameters associated with our solver are $C_{a_i}=0$ and $n_{lfi}=0,1$, whereas the panels (d)-(f) correspond to $C_{a_i}=1$ and $n_{lfi}=0,1$.

The initially circular droplet spreads and its final shape, depicted in Fig. \ref{figA1}, depends on the values of the surface tension coefficients of each system. It can be seen that the newly developed solver is capable of reproducing the correct trends in the droplet shape at the steady-state regime with respect to the overall surface tension effect, as indicated by the close agreement of our predictions with other published numerical results \cite{Xie2020, Kim2005}. This is valid for all combinations of the parameters $C_{a_i}$ and $n_{lfi}$ and it is also noted that our results appear to be in better agreement with the predictions of the more accurate scheme of Xie et al. \cite{Xie2020}. Turning our attention to the predictions of the original \texttt{multiphaseInterFoam} solver, also shown in Fig. \ref{figA1} for $C_{a_i}=0$ (see Fig. \ref{figA1}(a-c)) and $C_{a_i}=1$  (see Fig. \ref{figA1}(d-f)), we find large discrepancies relative to the both previously published results and our predictions, clearly pointing out a much lower accuracy of the original multiphase solver. These discrepancies appear to be unaffected by the inclusion of interface compression. A closer examination of Fig. \ref{figA1} also reveals that the latter predictions exhibit a very weak dependency on the surface tension coefficients, in clear disagreement with the rest numerical schemes. 

The interface compression parameter $C_{a_i}$ and the Laplacian filter parameter $n_{lfi}$ have a small but not negligible effect on the predicted droplet shape by our numerical scheme, as shown in Fig. \ref{figA1}. In particular, the application of filtering with $n_{lfi}=1$ leads to an increase of the drop's length accompanied by a reduction to the maximum height of the drop due to the conservation of its mass and area. This turns out to be more evident in the cases with $C_{a_i}=1$. On the other hand, the interface compression parameter seems to have exactly the opposite effect to that of the filtering operation. A reduction of the drop's length is apparent in Fig. \ref{figA1} when comparing cases without filtering $n_{lfi}=0$ having $C_{a_i}=0$ (upper row) with $C_{a_i}=1$ (lower row). The effect of these parameters on the solution accuracy will also be discussed in more detail below. 



\begin{figure}[t!]
\centerline{\includegraphics[width=0.5\linewidth] {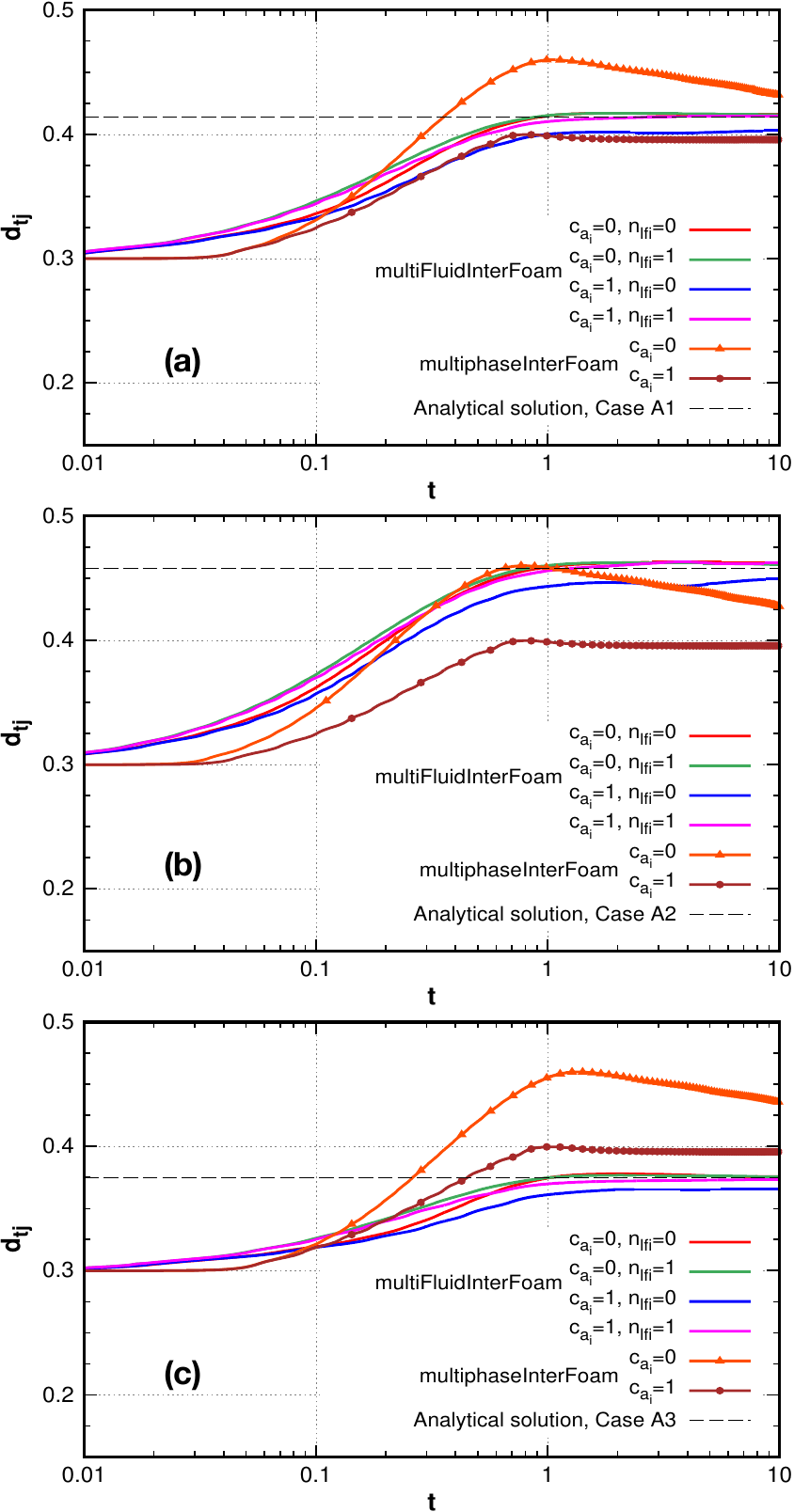}}
\caption{Time evolution of the interfaces of a floating lens for Case A$1$ with $We_{p,1}=We_{p,2}=We_{p,3}=60$ (a), Case A$2$ with  $We_{p,2}=60$ and $We_{p,1}=We_{p,3}=36$ (b), and Case A$3$ with $We_{p,2}=60$ and $We_{p,1}=We_{p,3}=108$ (c). The results are obtained for various values of $C_{a_i}$ and $n_{lfi}$ based on the new and the regular \texttt{multiphaseInterFoam} solvers. The dashed line depicts the equilibrium lens length (see Eq. \ref{eqDropAn1}). The density and viscosity ratios are $\rho_i=1$ and $\mu_i=1$, respectively.}
\label{figA4}
\end{figure}

The better performance of the new \texttt{multiFluidInterFoam} solver is also illustrated in Fig. \ref{figA4}, which shows the time evolution $t\,(=t^\ast \,U^\ast/L^\ast)$ of the non-dimensional lens' length $d_{tj}\,(=d_{tj}^\ast/L^\ast)$ for the same three cases (A1, A2 and A3). In this figure, we depict the prediction of our solver (solid lines) along with the results of the regular \texttt{multiphaseInterFoam} solver (solid lines with points). In addition, the dashed line depicts the analytical solution for the equilibrium lens' length; note that the derivation of an analytical solution for the equilibrium lens area, its length and the equilibrium contact angle is possible when the values of $\sigma_{ij}^\ast$ are specified. When a full drop is considered, the distance between the two triple junction points (denoted by $d_{tj}^\ast$ in Fig. \ref{figLensGeo}) can be calculated using the following expression \cite{Rowlinson2013, Liang2016, Yu2019}:

\begin{equation}\label{eqDropAn1}
d_{tj}^\ast=\sqrt{ \frac{4\,A^\ast}{\sum_{l=1}^2 \frac{1}{\sin\theta_l}(\frac{\theta_l}{\sin\theta_l}-\cos\theta_l)} } \, ,
\end{equation}

\noindent
where $A^\ast$ is the area of the droplet and $\theta_l \, (l = 1,2)$ is the contact angle as illustrated in Fig. \ref{figLensGeo}, determined as

\begin{equation}\label{eqDropAn2}
\cos{\theta_1}= \frac{\sigma_{13}^{\ast 2}+\sigma_{23}^{\ast 2}-\sigma_{12}^{\ast 2}}{2 \sigma_{23}^\ast \sigma_{13}^\ast} \,\,\, , \,\,\,\cos{\theta_2}= \frac{\sigma_{13}^{\ast 2}+\sigma_{12}^{\ast 2}-\sigma_{23}^{\ast 2}}{2 \sigma_{12}^\ast \sigma_{13}^\ast}\, .
\end{equation}

\noindent
After a transient time period, all numerical solutions provided by our solver approach the expected analytical solutions. Evidently, our solver predicts very well the correct long-term solution, for all cases (A1, A2, and A3) which correspond to different values of the interfacial tension coefficients. On the other hand, the original multiphase solver of \texttt{OpenFOAM} seems to be insensitive to the variation of the interfacial tension coefficients, showing almost the same transient behavior with varying $We_{p,1}$ and $We_{p,3}$ and providing an obviously wrong prediction for the long-term solution. 

Turning back our attention to the predictions of the present numerical scheme, we observe in Fig. \ref{figA4} that the predicted dynamics can be affected to some extent by the values of the $C_{a_i}$ and $n_{lfi}$ parameters and, therefore, their role and effect in the accuracy of the proposed solver should be discussed in more detail. In particular, we see in this figure that for $C_{a_i}=1$ and $n_{lfi}=0$ (i.e. with interface compression enabled and filtering disabled) the numerical scheme fails to predict the correct long term solution. This result provides an indication that the filtering of the volume fractions is essential in order to improve the quality of the predictions when interface compression is taken into account. Moreover, the time evolution of $d_{tj}$ when $C_{a_i}=0$ differs significantly from that when surface compression is enabled (with $C_{a_i}=1$), and the long-time solutions obtained in these cases do not coincide with each other. 
 

\begin{figure}[t!]
\centerline{\includegraphics[width=0.6\linewidth] {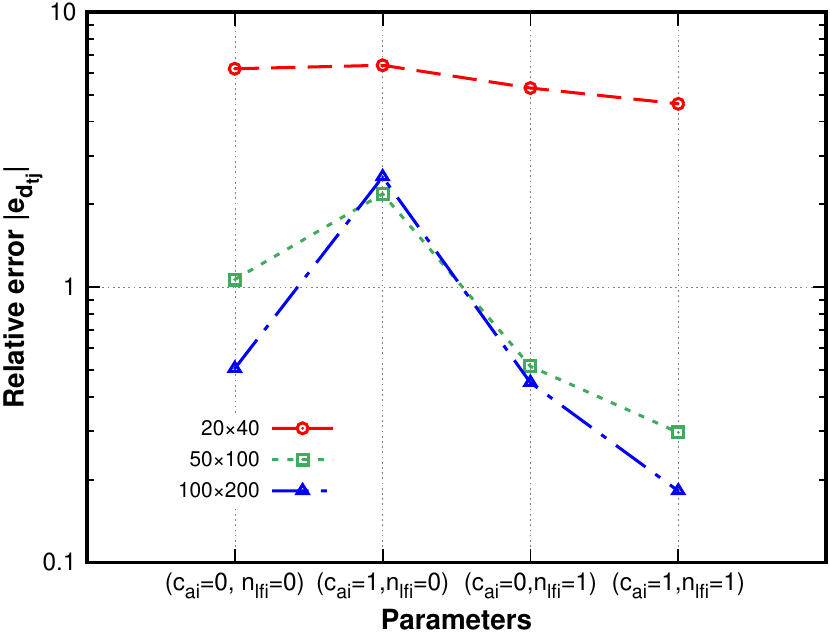}}
\caption{The relative error of the drop's length at equilibrium $e_{d_{tj}}$ between theory and numerical predictions for various combinations of the parameters $C_{a_i}$ and $n_{lfi}$ and three computational meshes consisting of $20 \times 40$, $50 \times 100$, and $100 \times 200$ control volumes. $We_{p,1}=We_{p,2}=We_{p,3}=60$, and the density and viscosity ratios are $\rho_i=1$ and $\mu_i=1$, respectively.}
\label{figA2}
\end{figure}

To obtain a clearer picture regarding the effect of the parameters $C_{a_i}$ and $n_{lfi}$ on the numerical prediction of the drop shape, the relative error of the drop's length was computed as $e_{d_{tj}}=(d_{tj,n}-d_{tj,a})/d_{tj,a}$ and it is shown in Fig. \ref{figA2}, where $d_{tj,n}$ and $d_{tj,a}$ are the values calculated numerically and analytically based on Eq. (\ref{eqDropAn1}) \cite{Rowlinson2013, Liang2016, Yu2019}, respectively. Results for three computational meshes of $20 \times 40$, $50 \times 100$, and $100 \times 200$ are presented. A number of observations can be made; at first, the relative error is reduced as the mesh is refined, with the exception of case ($C_{a_i}=1, \, n_{lfi}=0$), in which $|e_{d_{tj}}|$ seems to saturate at approximately $2\,\%$ as the grid spacing is reduced. In all the other cases, the value of $|e_{d_{tj}}|$ is below $1\,\%$. Secondly, it turns out that the adopted grid resolution of $100\times200$ is so fine and the time step is also small enough, so that the effect of spurious currents are rather insignificant and the results obtained with ($C_{a_i}=0$, $n_{lfi}=0$) are adequate. Thirdly, the improvement of the predictions when applying only filtering or when combining the sharpening and filtering operations is more evident for the smaller mesh of $20\times40$ control volumes. The absolute error difference $\Delta (e_{d_{tj}})^{C_{a_i},n_{lfi}}= (e_{d_{tj}})^{C_{a_i},n_{lfi}} - (e_{d_{tj}})^{0,0}$ can be used for quantification purposes. For example, it is $\Delta (e_{d_{tj}})^{0,1}$= 0.919, 0.551, and 0.056 for the meshes $20 \times 40$, $50 \times 100$, and $100 \times 200$, respectively, while it holds that $\Delta (e_{d_{tj}})^{1,1}$= 1.579 ($20 \times 40$), 0.77 ($50 \times 100$), and 0.324 ($100 \times 200$). It is worth mentioning that the opposite is observed for the effect of the $C_{a_i}$ parameter, as it is $\Delta (e_{d_{tj}})^{1,0}$=0.188, 1.114, and 2 for non-dimensional grid spacings of $\Delta x=1/40$, $1/100$, and $1/200$, respectively. This points out that the optimum $C_{a_i}$ depends on the grid resolution, revealing that there is no need for application of excessive levels of artificial sharpening of interfaces since the interfaces are well captured as the grid is refined. The above discussion suggests a stronger impact of the filtering and combined filtering/sharpening operations in coarser simulations. Finally, the worst and best performances are found for the parameter combination ($C_{a_i}=1$, \, $n_{lfi}=0$) and ($C_{a_i}=1$, \, $n_{lfi}=1$), respectively. A better agreement of the present numerical solutions with the analytical solutions and previous data of Xie et al. \cite{Xie2020} and Kim and Lowengrub \cite{Kim2005} is generally observed when using the new solver with $C_{a_i}=1$ and $n_{lfi}=1$. 

\begin{center}
\begin{table*}[t!]%
\caption{A comparison of the non-dimensional equilibrium distance between triple junctions $d_{tj}$ in three representative cases obtained from analytical solutions, the present solver ($C_{a_i}=1$, $n_{lfi}=1$, $\Delta x=1/200$), the original \texttt{OpenFOAM} solver ($C_{a_i}=1$, $\Delta x=1/200$), and past works with $\Delta x=1/256$ \cite{Xie2020, Kim2005}. The relative error of each numerical prediction against the analytical solution is shown in the parentheses. \label{tab1}}
\centering
\begin{adjustbox}{max width=1\textwidth}
\begin{tabular*}{650pt}{@{\extracolsep\fill}lcccccccc@{\extracolsep\fill}}
\toprule
\textbf{Case} & $We_{p,1}$  & $We_{p,2}$  &  $We_{p,3}$   &  \textbf{Analytical}  & \textbf{Present solver}  & \textbf{OpenFOAM solver}  & \textbf{Xie et al. \cite{Xie2020}}  &   \textbf{Kim and Lowengrub \cite{Kim2005}}   \\
\midrule
A1 & 60    &   60  &  60    &   0.4138  &   0.4145 (0.17\%)   &   0.3957 (4.37\%)   &   0.4069 (1.67\%)   &   0.4368 (5.56\%) \\
A2 & 36   &   60  &  36   &   0.4578  &   0.4626  (1.05\%)  &   0.3958 (13.54\%)   &   0.4485 (2.03\%)   &   0.4622 (0.96\%) \\
A3 & 108  &   60  &  108  &   0.3746  &   0.3735 (0.29\%)   &   0.3957 (5.63\%)   &   0.3701 (1.2\%)   &   0.3982 (6.3\%) \\
\bottomrule
\end{tabular*}
\end{adjustbox}
\end{table*}
\end{center}

In Table \ref{tab1}, we compare the length $d_{tj}$ predicted by the numerical simulations and compare it against both the analytical solutions and two previous works \cite{Xie2020, Kim2005}. The results of the new solver compares favourably with both the analytical solution and the results of Xie et al. \cite{Xie2020} and Kim and Lowengrub \cite{Kim2005}. A quantitative comparison with the analytical solution reveals that the relative error of the \texttt{multiFluidInterFoam} solver is $0.18 \%$, $1 \%$, and $0.28 \%$ for Cases A$1$, A$2$, and A$3$ respectively, providing further confirmation of the excellent accuracy of the present solver. On the other hand, using the exact same grid of $100 \times 200$ control volumes, the relative error for the original \texttt{multiphaseInterFoam} solver is $4.7 \%$ (Case A$1$), $13.54 \%$ (Case A$2$), and $5.62 \%$ (Case A$3$). Our solver achieves the same or better level of accuracy with respect to the methods of Xie et al. \cite{Xie2020} and Kim and Lowengrub \cite{Kim2005}, as also indicated in Table \ref{tab1}. It is pointed out that the present results are obtained with the use of a slightly coarser non-dimensional grid spacing of $\Delta x=1/200$ as compared with $\Delta x=1/256$ adopted in the aforementioned works.

\begin{figure}[t!]
\centerline{\includegraphics[width=0.65\linewidth] {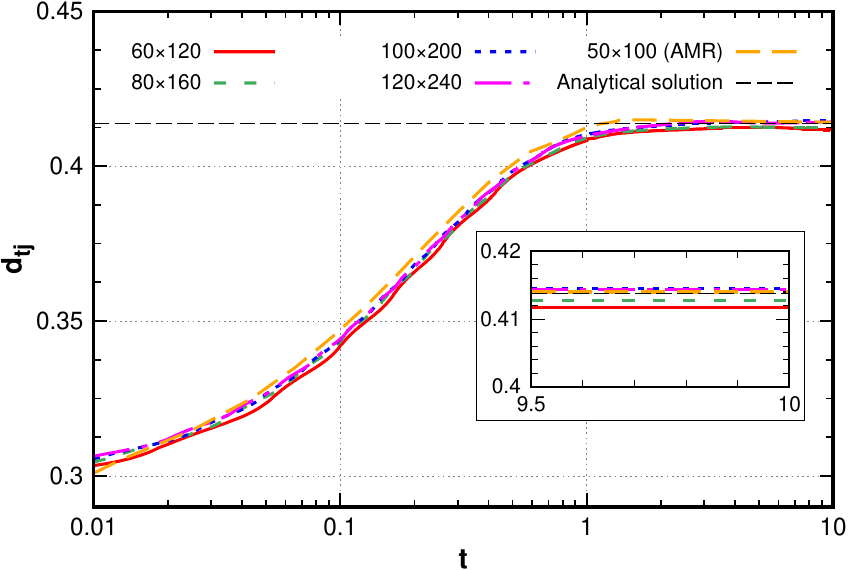}}
\caption{Time evolution of the interfaces of a floating lens for $We_{p,1}=We_{p,2}=We_{p,3}=60$ for various fixed uniform computational meshes $60 \times 120$, $80 \times 160$, $100 \times 200$, $120 \times 240$, and an adaptive mesh of $50 \times 100$ with a minimum non-dimensional grid spacing of $1/200$. The density and viscosity ratios are $\rho_i=1$ and $\mu_i=1$, respectively.} 
\label{figA5}
\end{figure}

Finally, in order to assess how the predictions of the new solver are affected by the selected grid, a convergence study is presented below. Beside the standard adaptive mesh refinement algorithm provided by \texttt{OpenFOAM}, the present solver also employs a proper adaptation of this mesh tool for two dimensional problems \cite{AMR2d}. Therefore, we performed simulations using four static meshes and one simulation which was carried out with an adaptive mesh for the representative Case A$1$ (i.e. for $We_{p,1}=We_{p,2}=We_{p,3}=60$). Fig. \ref{figA5} shows the comparison between the numerical simulations with the different grids for the evolution of the droplet length, $d_{tj}$, while the analytical solution at equilibrium is depicted with the dashed line. Sharpening of interfaces is activated with $C_{a_i}=1$ and smoothing operation in the calculation of curvature is applied with $n_{lfi}=1$. The time step in these simulations is set to $\Delta t^\ast=10^{-7}\, \mathrm{s}$. It can be easily concluded that the results shown here are converged. More specifically, the long-term solution of $d_{tj}$ is $0.4117$, $0.4127$, $0.4145$, $0,4143$ for the fixed meshes of $60 \times 120$ ($1$), $80 \times 160$ ($2$), $100 \times 200$ ($3$), and $120 \times 240$ ($4$), respectively, and $0.414$ for the adaptive mesh of $50 \times 100$ with a minimum non-dimensional grid spacing of $1/200$. The numerical errors for the different meshes are $0.5\%$, $0.26\%$, $0.18\%$, $0.12\%$ for the fixed mesh $1$, $2$, $3$, and $4$, respectively, and $0.05\%$ for the adaptive mesh. It is noted that, for the adaptive mesh, the total number of control volumes needed to achieve the same level of accuracy is considerably smaller as the mesh is only refined near the interfaces. The grid is locally refined in regions in which $0.01 \le \sum_i a_i \le 0.99$ every $10$ time steps ($\textit{refineInterval}$). Both the maximum refinement level ($\textit{maxRefinement}$) and the buffer layers ($\textit{nBufferLayers}$) are set to unity. Converged results and similar findings are observed by varying the values of $We_{p,1}$, $We_{p,3}$, $C_{a_i}$ and $n_{lfi}$, and they are not shown here.

\subsubsection*{\it Full spreading} \label{sec3.3.4}

We then consider the case of complete spreading, where the three interfacial tensions cannot yield a Neumann's triangle. The parameters for the simulated case of complete spreading are $\sigma_{13} = 10$, and $\sigma_{12}=\sigma_{23} = 1$.  Clearly, it holds that $S_2=\sigma_{13} - \sigma_{12} - \sigma_{23}>0$, which corresponds to a supercritical state. All the other numerical details and parameters are those given in Section \ref{sec3.3.2}.

\begin{figure}[h!]
\centerline{\includegraphics[width=0.97\linewidth] {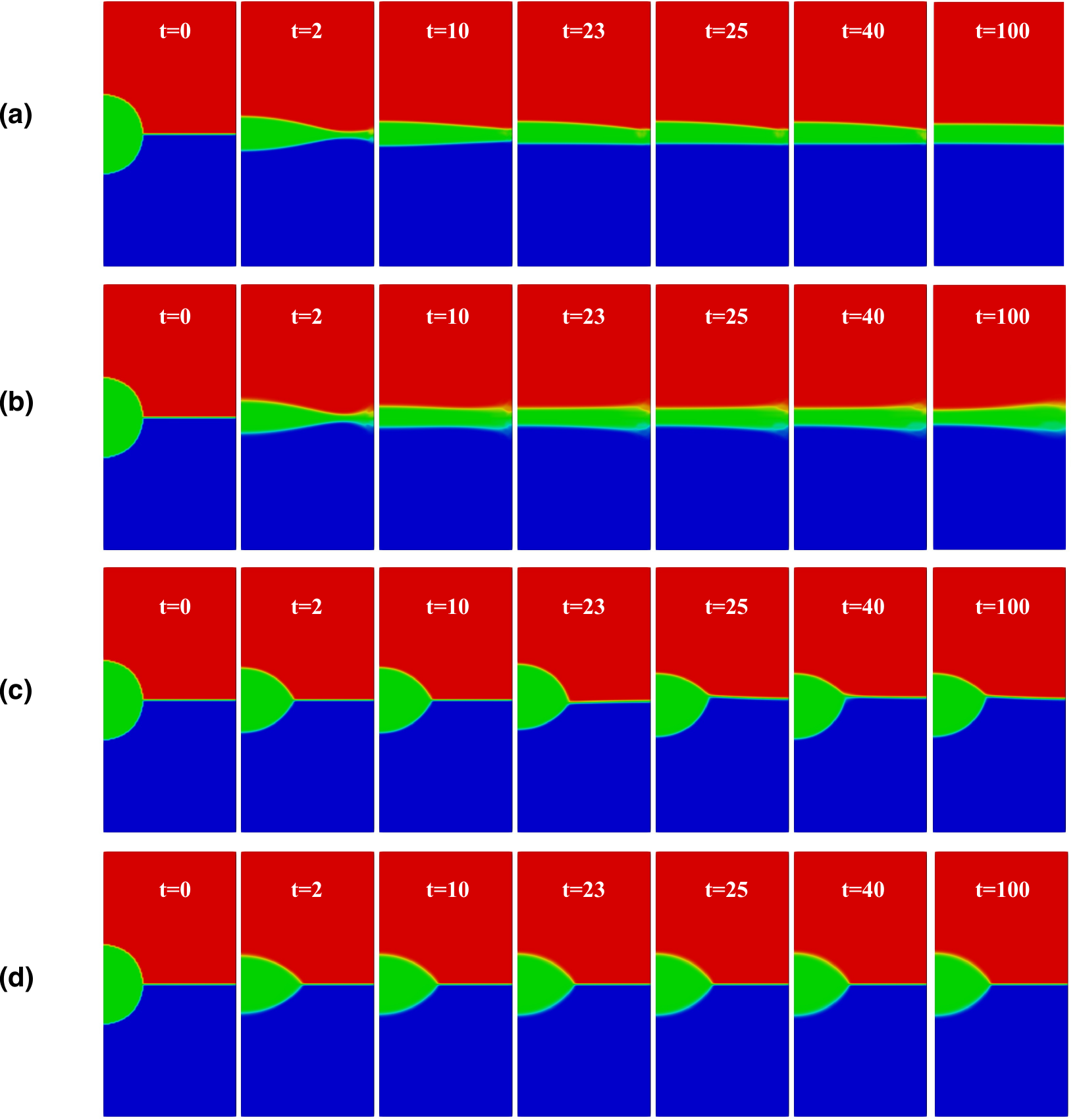}}
\caption{Complete spreading of an initially spherical droplet under surface tension forces at different time instants. Results are obtained based on the new \texttt{multiFluidInterFoam} solver with $C_{a_i}=1$, $n_{lfi}=1$  (a) and $C_{a_i}=0$, $n_{lfi}=0$  (b), and by using the original \texttt{multiphaseInterFoam} solver with $C_{a_i}=1$ (c) and $C_{a_i}=0$ (d). The density and viscosity ratios are $\rho_i=1$, $\mu_i=1$, respectively, while the surface tension coefficient ratios are $\sigma_{13}=10$, $\sigma_{12}=\sigma_{23}=1$.}
\label{figA3}
\end{figure}

A complete spreading of phase $2$ (green) is anticipated, which is closely reproduced by the new multiphase solver with $C_{a_i}=1$, $n_{lfi}=1$ and $C_{a_i}=0$, $n_{lfi}=0$ as seen in Figs. \ref{figA3}(a) and (b), respectively. The initially circular droplet evolves finally into a plane fluid layer located within the upper and lower fluids. An obvious blurring occurs near the interfaces when the artificial compression is deactivated, whereas sharper interfaces can be observed for $C_{a_i}=1$. The size of the computational domain is small and the right wall is close to the droplet. When fluid $2$ hits the right wall, it has sufficient momentum to bounce back, forming eventually a wave. It is found that this wave takes a long time period to be suppressed. An even longer time is required when interface compression and smoothing of the volume fractions are not considered. 

Fig. \ref{figA3} also shows the time evolution of the droplet shape and interfaces predicted by using the original \texttt{multiphaseInterFoam} solver with $C_{a_i}=1$ and $0$. In both cases, the green droplet keeps adhering to the interface between the upper and lower fluids and three-phase regions are clearly seen at long times. Since the three interfacial tensions do not satisfy a Neumann's triangle, a stable three-phase region cannot be formed and the results shown in Figs. \ref{figA3}(c) and (d) are clearly not correct. The interface sharpening, as implemented in \texttt{multiphaseInterFoam}, introduces also instabilities in the numerical solutions as indicated by the up and down movements of the drop at times $t=23$ to $25$, which affect the shape of the drop. 

\subsubsection{Drop levitation} \label{sec3.3.5}
In this Section, we consider a droplet leaving an interface under surface tension forces \cite{Smith2002, Xie2020, Tofighi2013}. Based on the aforementioned studies, we use the same computational setup as in the floating lens problem in Section \ref{sec3.3.2}, where a circular droplet (fluid $2$, green) is initialized between the interface of fluid $1$ (blue) and fluid $3$ (red) with zero velocity. Here, we consider a droplet levitation problem by setting the interfacial tension between top and bottom fluids $\sigma_{13}$ and the interfacial tension between droplet and top fluid $\sigma_{23}$ equal to unity, whereas the interfacial tension between droplet and bottom fluid $\sigma_{12}$ is set to $10$. The spreading parameter is positive $S_3=\sigma_{12}-\sigma_{13}-\sigma_{23}>0$, indicating a complete spreading of the top fluid along the droplet in such a way that the top fluid becomes entrained between the droplet and its subphase and the triple points disappear. 

\begin{figure}[h!]
\centerline{\includegraphics[width=1\linewidth] {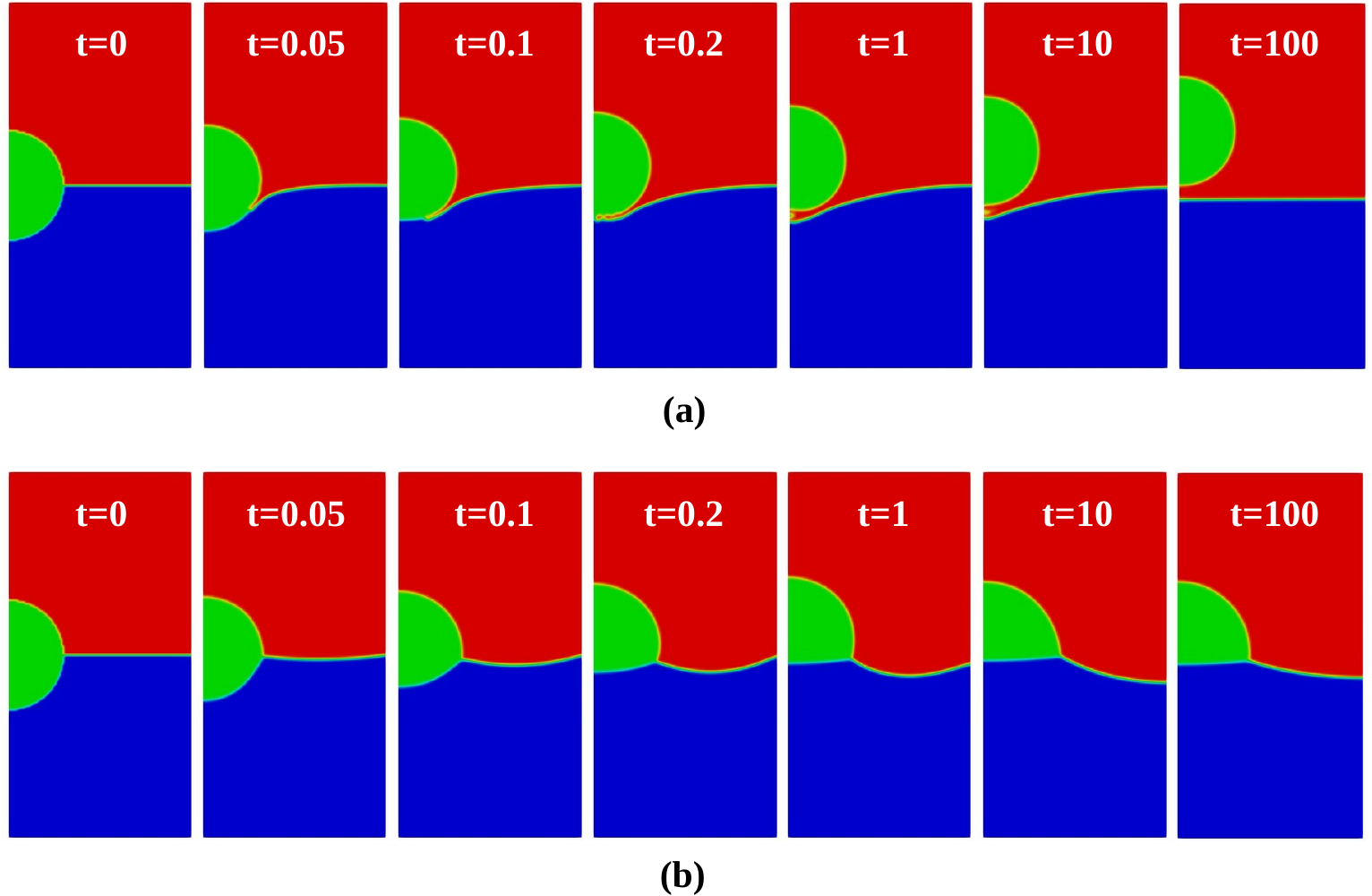}}
\caption{Evolution of a droplet leaving an interface under the action of surface tension forces a different time instants. Here, the density and viscosity ratios of the phases are  $\rho_i=1$ and $\mu_i=1$, respectively, and $\sigma_{12}=10$, $\sigma_{13}=\sigma_{23}=1$. Results are obtained by the new \texttt{multiFluidInterFoam} solver with $C_{a_i}=1$, $n_{lfi}=1$ (upper row) and the regular \texttt{multiphaseInterFoam} solver with $C_{a_i}=1$ (lower row).}
\label{figC1}
\end{figure}

Fig. \ref{figC1}(a) shows the predicted results for the evolution of the droplet due to the flow caused by the balance of the surface tension forces. The new multiphase solver with $C_{a_i}=1$ and $n_{lfi}=1$ is able to deal with large deformation of the interface, yielding results similar to the those of Smith et al. \cite{Smith2002}, Xie et al. \cite{Xie2020} and Tofighi et al. \cite{Tofighi2013}. At large times, the whole droplet resides completely inside the upper fluid. Consistently with what is expected from theory, the system tends to an equilibrium state in which the interfaces are finally plane or spherical. The total area for the droplet for fluid $2$ is kept constant during the simulation pointing out the conservative nature of the new multiphase VOF based solver.

In contrast, the original \texttt{multiphaseInterFoam} solver with $C_{a_i}=1$ cannot reproduce the levitation of the droplet and it predicts a steady shape of the lens at long simulation times. Similar results are obtained when the surface compression is deactivated. It is recalled that the combination of surface tension coefficients in this Section do not satisfy a Neumann's triangle and, thus, stable-three phase regions cannot be formed. Consequently, the results produced by the original multiphase solver of \texttt{OpenFOAM} contradict theoretical anticipations and they are obviously incorrect.  

\subsection{Consistency and verification tests for the \texttt{rheoMultiFluidInterFoam} solver} \label{sec3.2}

We now proceed with the verification of the second multiphase solver \texttt{rheoMultiFluidInterFoam}. Since there is no rigorous benchmark for non-Newtonian three-phase fluid flows, for the validation of our solver we will examine the two-phase problem of a bubble rising both in a viscoplastic (described by the Herschel--Bulkley model) and a viscoelastic (described by the Phan-Thien Tanner model) fluid. The latter serves as a reference to validate the presently \texttt{RheoTool} based implementation of the model and check the consistency of our solver in the limiting case of two-phase flows. Finally, we proceed with the simulation of typical three-phase flows, such as the floating lens and levitating drop problems, which are re-examined by assuming that the bottom fluid layer exhibits a viscoplastic behavior (described by the Herschel--Bulkley model), while the predictions of our generic Non-Newtonian multiphase solver are compared against the previously presented \texttt{multiFluidInterFoam} solver, which provides a different implementation of the Herschel--Bulkley model.

\subsubsection{Rising bubble in a viscoplastic fluid} \label{sec3.2.1}

The buoyancy-driven rise of an axisymmetric bubble in a viscoplastic material has been studied in detail \cite{Dimakopoulos2013, Tsamopoulos2008, Karapetsas2019} and we have chosen this problem to validate the second multiphase solver \texttt{rheoMultiFluidInterFoam} in the limiting case of two-phase flow involving a non-Newtonian liquid described by a Generalized Newtonian model, such as the regularized Herschel--Bulkley model.

\begin{figure}[h!]
\centerline{\includegraphics[width=0.425\linewidth] {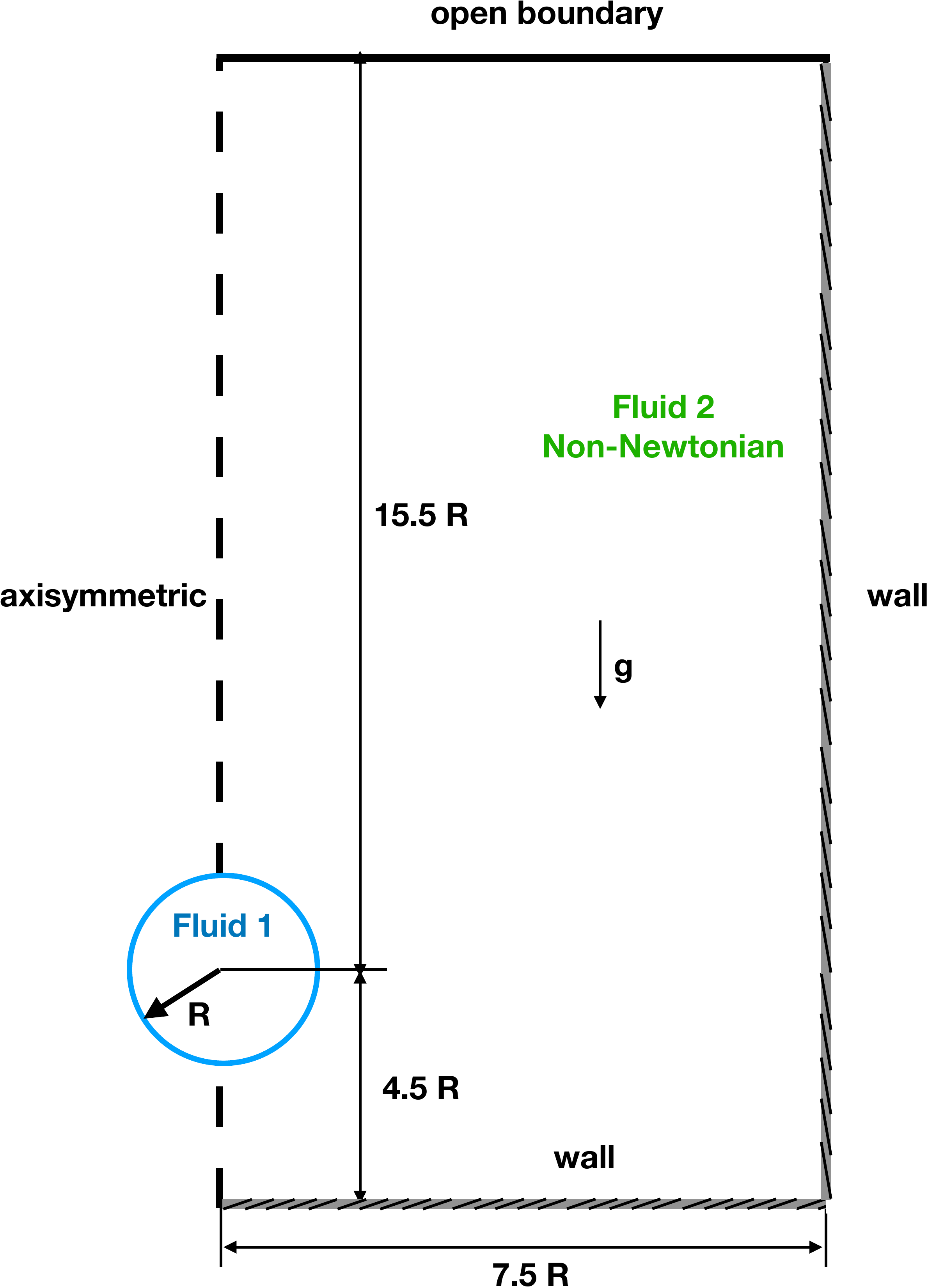}}
\caption{Geometry and configuration of a bubble rising in a viscoplastic fluid.}
\label{D1}
\end{figure}

Initially, a spherical bubble with a non-dimensional radius $R_b$ is placed at a distance of $4.5 R_b$ from the bottom rigid wall. The bubble moves upward opposed to the action of gravity in a viscoplastic fluid. Fig. \ref{D1} shows the geometry and the initial configuration of the rising bubble problem. Detailed simulations of a bubble rising in Newtonian or viscoplatic materials following the Papanastasiou regularization \cite{Papanastasiou1987} were undertaken by Tsamopoulos et al. \cite{Tsamopoulos2008}. These are used as a reference for comparison against our numerical predictions. Both the bounded and the Papanastasiou regularization versions of the Hesrchel--Bulkley model are utilized here. It is recalled that the former implentation is the only one available in the \texttt{multiFluidInterFoam} solver, whereas both regularization methods are implemented in the \texttt{rheoMultiFluidInterFoam} solver based on the \texttt{RheoTool} toolbox.   

The non-dimensional parameters characterizing this system are: the Bond number $Bo=\rho_p^\ast g^\ast R_b^{\ast 2} / \sigma_{bp}^\ast$, the Archimedes number $Ar=\rho_p^{\ast 2} g^\ast R_b^{\ast 3} / \mu_p^{\ast\,2}$, and the Bingham number $Bn=\tau_0^\ast / (\rho_p^\ast g^\ast R_b^\ast)$. Here, $R_b^\ast$ denotes the radius of the bubble, $\rho_b^\ast$, $\mu_b^\ast$ are the density and dynamic viscosity of the bubble, respectively, and $\sigma_{bp}^\ast$ is the surface tension coefficient between the two phases. The density of the pseudo-plastic fluid and its dynamic viscosity are $\rho_p^\ast$ and $\mu_p^\ast$, respectively, $\tau_0^\ast$ is the yield stress of the material, and $g^\ast$ is the gravity acceleration. Proper scales for the length, velocity, and stresses are $R_b^\ast$, $\rho^\ast g^\ast R_b^{\ast \, 2}/\mu_p^\ast$, and $\rho^\ast g^\ast R_b^\ast$, respectively. The parameters used in the study are $Ar=50$, $Bo=10$, and $Bn=0$ and $0.14$; these dimensionless parameters correspond to a system with $\rho_p^\ast=1000 \, \mathrm{Kg}\,\mathrm{m}^{-3}$, $\rho_b^\ast=1 \, \mathrm{Kg}\,\mathrm{m}^{-3}$, $\mu_p^\ast=1.414\times10^{-2} \, \mathrm{Kg}\,\mathrm{m}^{-1}\mathrm{s}^{-1}$, $\mu_b^\ast=10^{-5} \, \mathrm{Kg}\,\mathrm{m}^{-1}\mathrm{s}^{-1}$, $\tau_0^\ast=1.4\,\mathrm{Pa}$, $R_b^\ast=0.001\,\mathrm{m}$, and $g^\ast=10 \, \mathrm{m}\mathrm{s}^{-2}$.


No-slip boundary condition is applied to the walls, while the top surface is assumed to be an open boundary. Zero gradient was enforced at the walls for the bubble volume fraction, while at the open boundary it was set to zero. Axial symmetry is applied for all variables at the left boundary. The size of the computational domain is $7.5R_b \times 20R_b$ in the $r$, $z$ directions, respectively. The resolution of the computation mesh is $150 \times 400$ control volumes. The time step is determined by using a global and local Courant numbers of $0.05$, while the maximum time step allowed is set equal to $\Delta t^\ast=10^{-5}\,\mathrm{s}$. Interface compression is taken into account by setting the parameter $C_{a}$ to unity, while no filtering is applied. All the results shown in this Section are obtained based on the \texttt{rheoMultiFluidInterFoam} solver. The same case could be easily examined by using the \texttt{rheoInterFoam} solver which is the \texttt{RheoTool} counterpart of the standard \texttt{interFoam} solver and it can handle two-phase flows. It was verified that both solvers produce the same results in the case under investigation (not presented herein for conciceness), without considerable differences in the execution times.

\begin{figure}[t!]
\centerline{\includegraphics[width=0.75\linewidth] {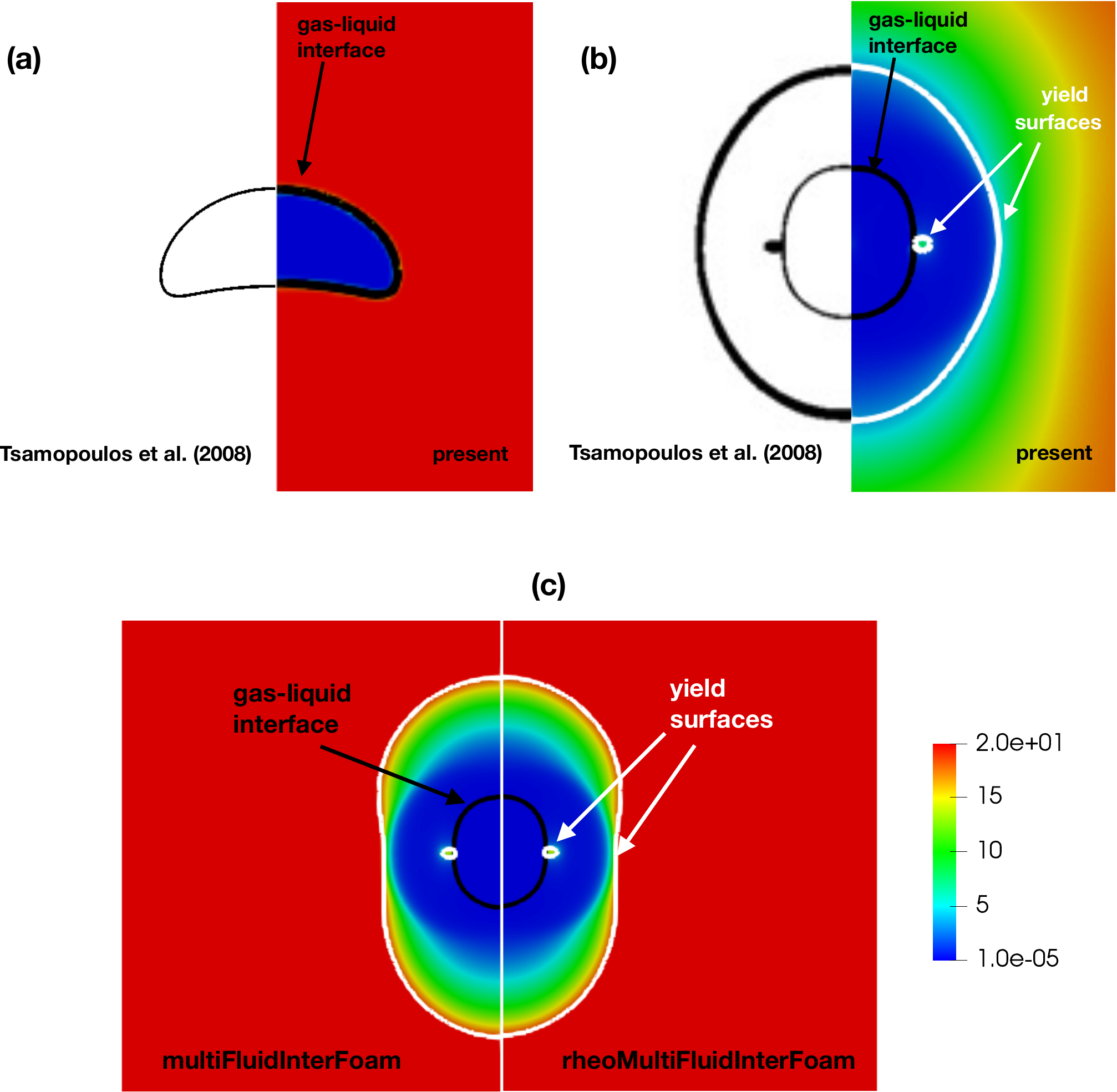}} 
\caption{(a) Bubble shape (black line) in a Newtonian fluid for $Ar=50$ and $Bo=10$ and (b) bubble shape (black line) and $|\mathbf{\underline{\tau}}|=Bn$ isoline (white line) in a viscoplastic fluid for $Ar=50$ and $Bo=10$, $Bn=0.14$, and $N=10^4$, predicted in the present work by the Herschel--Bulkley model with the Papanastasiou regularization (right half) and in Tsamopoulos et al. \cite{Tsamopoulos2008} (left half). (c) Comparison of bubble shape (black line), $|\mathbf{\underline{\tau}}|=Bn$ isoline (white line) and dynamic viscosity obtained by the \texttt{multiFluidInterFoam} (left half) and \texttt{rheoMultiFluidInterFoam}  (right half) solvers together with the bounded bi-viscosity Herschel--Bulkley model. The parameters are the same as in (b).}
\label{D2}
\end{figure}

Fig. \ref{D2}(a) shows results regarding the shape of the bubble for a representative Newtonian case ($Bn=0$, $n=1$) with $Ar=50$ and $Bo=10$ for which a steady-state solution is generally expected. It can be seen that the gas bubble changes from its initially spherical to a steady oblate-spheroid shape. We find excellent agreement when comparing our bubble shape with that predicted by Tsamopoulos et al. \cite{Tsamopoulos2008}, shown in left half of Fig. \ref{D2}(a).

The case of a viscoplastic surrounding fluid is examined in Fig. \ref{D2}(b) where we depict the bubble shape obtained in steady-state for a characteristic case with $Ar=10$, $Bo=50$, along with a color map of the dynamic viscosity Here we have considered the case for $Bn=0.14$ and $n=1$ while employing the Papanastasiou regularization model for $N=m^\ast \, \rho_p^\ast g^\ast R_b^\ast / \mu_p^\ast=10^4$. As seen in Fig. \ref{D2}(b), the bubble takes a bullet-like shape due to the fact that the effective viscosity around the equatorial plane is higher than that at the poles. To determine the size and shape of the yield/unyieled regions, there are two criteria that have been employed by several researchers in the past. The location of the yield surface can be defined as the locus where either: a) $\dot{\gamma}^\ast=0$, or b) $|\mathbf{\underline{\tau}}^\ast|=\tau_0^\ast$. Although these criteria are equivalent in principle according to the discontinuous Bingham model, they are not equivalent when a regularized model such as the Papanastasiou or the bi-viscosity model are used \cite{Tsamopoulos2008}. Only the second criterion is appropriate for regularized models, i.e. that the material flows when the second invariant of the extra stress tensor exceeds the yield stress, and has been used for the results presented herein. This criterion in its dimensionless form becomes $|\mathbf{\underline{\tau}}|> Bn$ for yielded material and $|\mathbf{\underline{\tau}}| \le Bn$ for unyielded material. In Fig. \ref{D2}(b), the yield surface (isoline of $|\mathbf{\underline{\tau}}|=Bn$ shown with white line) is also plotted revealing that two regions of unyielded material appear, where the rate of strain is low enough around so that the experienced stresses are below the yield stress of the material: one far from the bubble and another around the equatorial plane, in excellent agreement with the results of Tsamopoulos et al. \cite{Tsamopoulos2008}.

In Fig. \ref{D2}(c) we provide a third consistency test by comparing the predictions between the \texttt{multiFluidInterFoam} and \texttt{rheoMultiFluidInterFoam} when the bounded bi-viscosity model is used which is the only available regularization method in the former solver. To provide as accurate predictions as possible with the bi-viscosity model we have set the upper bounding value of the viscosity, $\mu_{0,p}^\ast$, to an appropriate value to maintain the same level of accuracy with the Papanastasiou model presented above; the latter is achieved by setting $\mu_{0,p}^\ast = (1 + Bn \, N) \mu_p^\ast$. As seen in Fig. \ref{D2}(c) the two solvers provide nearly identical predictions.

Finally, at this point it also is appropriate to make a comment on the use of the \texttt{PIMPLE} algorithm for the velocity-pressure coupling which is regularly used in the original \texttt{multiphaseInterFoam} solver. The transient simulations presented in this Section using \texttt{rheoMultiFluidInterFoam} solver employ the \texttt{SIMPLEC} algorithm, as mentioned in Section \ref{sec2.2.4}. It was found that when using the standard \texttt{PIMPLE} algorithm, a sufficiently smaller time step $\Delta t^\ast$ was required or a greater number of cycles were necessary in order to obtain meaningful and correct solutions. The same behavior was also found when using the \texttt{multiphaseInterFoam} solver which employs the \texttt{PIMPLE} method. To assess the effect of the velocity-pressure coupling we summarize in Table \ref{tab2}, five different numerical experiments for $Ar=50$, $Bo=10$, $n=1$, $Bn=0.14$; the Papanastasiou model has been employed with $N=10^4$. In case B1, the \texttt{SIMPLEC} algorithm exhibiting by far the best behaviour in terms of computational time. On the other hand, when the \texttt{PIMPLE} algorithm is used, e.g. see cases B2 and B3, where a higher number of inner iterations $n_{ii}$ of the coupling method was used while maintaining the same time step, an unsteady flow was predicted due to numerical instabilities; the high viscosity of the viscoplastic material in the unyielded regions generated numerical instabilities near the walls that traveled toward the inner region. A careful inspection, though, of the local viscosity revealed that this behaviour was merely a numerical artifact, which significantly deteriorated the accuracy of the solver. By either significantly increasing the number of inner iterations (case B4) or decreasing the time step, $\Delta t^\ast$, (case B5) the correct steady flow regime was predicted, albeit with considerably increased computational cost.

\begin{center}
\begin{table*}[t]%

\caption{Summary of the execution time and the resulting flow regime obtained by the \texttt{rheoMultiFluidInterFoam} solver when using the \texttt{SIMPLEC} and the \texttt{PIMPLE} algorithms for various time steps $\Delta t^\ast$ and inner iterations $n_{ii}$ of the velocity-pressure coupling methods. These simulations were performed in parallel using $4$ cores for a total simulation time of $T^\ast_{sim}=0.001\,\mathrm{s}$.  The parameters are $Ar=50$, $Bo=10$, $n=1$, $Bn=0.14$; the Papanastasiou model has been employed with $N=10^4$.}\label{tab2}
\centering
\begin{adjustbox}{max width=\textwidth}
\begin{tabular*}{500pt}{@{\extracolsep\fill}lccccc@{\extracolsep\fill}}
\toprule
\textbf{Case} & \textbf{Velocity-pressure coupling} & $\mathbf{\Delta t}^\ast$ \textbf{[s]}  &  $\textbf{n}_{ii}$   & \textbf{Flow regime}  &  \textbf{Execution time [s]} \\
\midrule
B$1$ & \texttt{SIMPLEC}  & $10^{-5}$    &  1      &   steady-state    &   231.2    \\
B$2$ & \texttt{PIMPLE}    & $10^{-5}$    &  3      &   unsteady         &   323.8     \\
B$3$ & \texttt{PIMPLE}    & $10^{-5}$    &  10    &   unsteady         &   614.3     \\
B$4$ & \texttt{PIMPLE}    & $10^{-5}$    &  50    &   steady-state    &   1714.1   \\
B$5$ & \texttt{PIMPLE}    & $10^{-6}$    &  3      &   steady-state    &   2566.3   \\
\bottomrule
\end{tabular*}
\end{adjustbox}
\end{table*}
\end{center}

\subsubsection{Rising droplet in a viscoelastic fluid} \label{sec3.2.2}

In this Section, the \texttt{rheoMultiFluidInterFoam} solver is validated for a Newtonian droplet rising in a viscoelastic fluid described by the Phan-Thien Tanner model. We consider a two-dimensional (2D) rectangular domain having width $L_x^\ast$ and height $L_y^\ast$. Initially, a spherical droplet with a radius $R_{d,0}^\ast$ is located at ($L_x^\ast/2$, $L_y^\ast/4$). The no-slip condition is applied at the horizontal boundaries, whereas the free-slip condition is applied at the vertical boundaries. The viscosity $\mu_2^\ast$ and density $\rho_2^\ast$ of droplet fluid 2 are smaller than those $\mu_1^\ast$, $\rho_1^\ast$ of the surrounding fluid 1; $\mu_1^\ast = \mu_{s,1}^\ast+\mu_{p,1}^\ast$, where $\mu_{s,1}^\ast$ and $\mu_{p,1}^\ast$ denote the solvent and the polymeric viscosities of the viscoelastic fluid. The gravity vector $\mathbf{g}^\ast$ points toward the bottom of the domain. Proper reference scales are $L_{ch}^\ast = 2 R_{d,0}^\ast$, $U_{ch}^\ast = \sqrt{g^\ast L_{ch}^\ast}$, $L_{ch}^\ast / U_{ch}^\ast$ for the length, velocity, and time, respectively. The same case has been examined by Prieto \cite{Prieto2015}.The relevant non-dimensional parameters are: the Reynolds number $Re = \rho_1^\ast U_{ch}^\ast L_{ch}^\ast / \mu_1^\ast$, the Weber number $We = \rho_1^\ast U_{ch}^{\ast,2} L_{ch}^\ast / \sigma_{12}^\ast$, the Weissenberg number $Wi = \lambda^\ast U_{ch}^\ast / L_{ch}^\ast$, the viscosity ratio $r_\mu = \mu_2^\ast / \mu_1^\ast$, the density ratio $r_\rho = \rho_2^\ast / \rho_1^\ast$, and the solvent viscosity ratio $\beta_1=\mu_{s,1}^\ast/\mu_1$. The quantity $\sigma_{12}^\ast$ is the surface tension coefficient of the interface between the two fluids. The linear PTT model with $\epsilon_1=\zeta_1=0$ is used here to facilitate a direct comparison with the results of Prieto \cite{Prieto2015}, who utilised the microscopic Hooke model. 

\begin{figure}[t!]
\centerline{\includegraphics[width=0.9\linewidth] {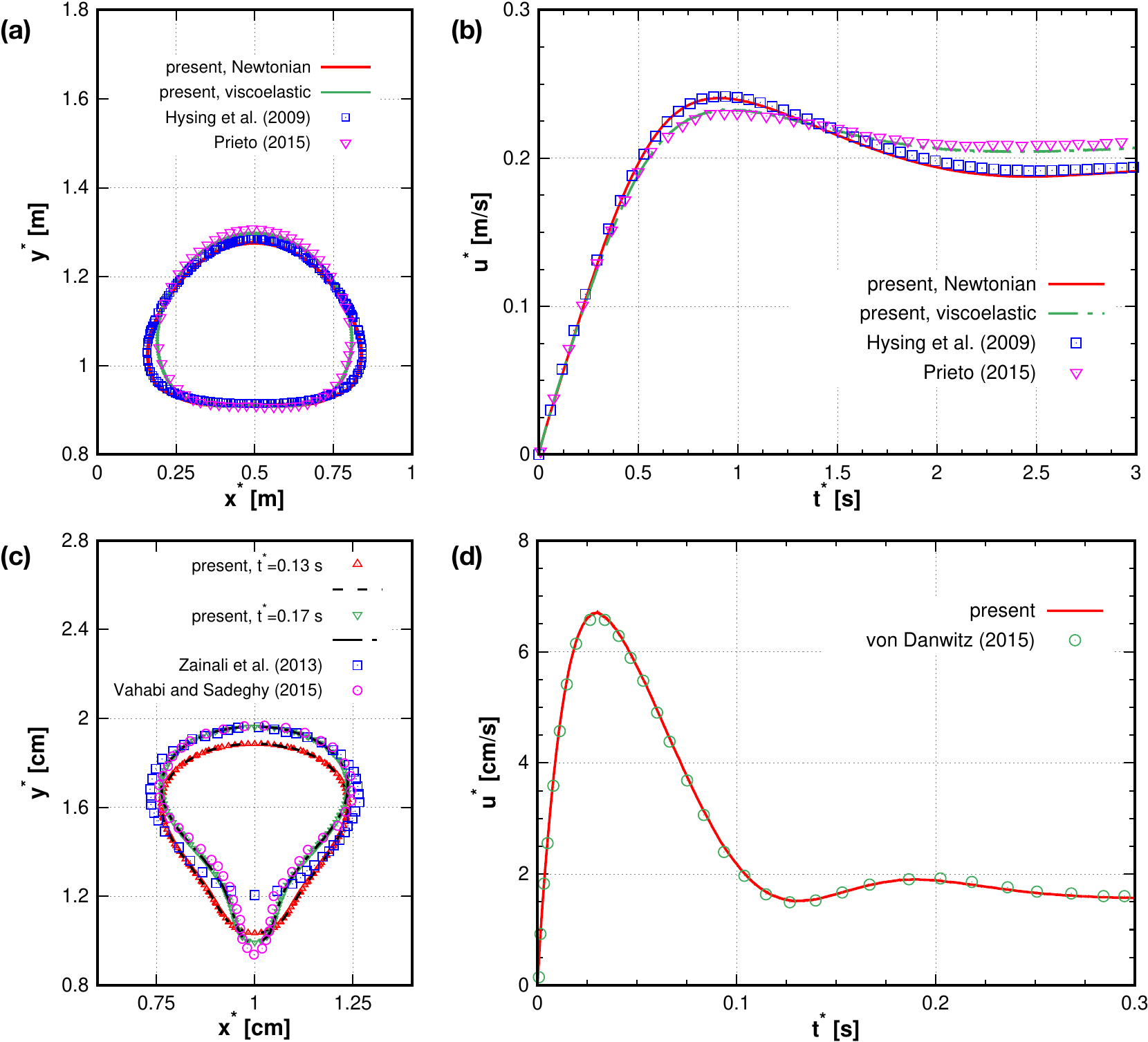}}
\caption{(a) Steady-state shape and (b) time evolution of the rise velocity for a Newtonian droplet in Newtonian (Case NN) and viscoelastic (Case NV1) mediums. The lines denote results obtained by our solver, while symbols correspond to previous data of Hysing et al. \cite{Hysing2009} and Prieto. \cite{Prieto2015}. (c) Droplet shape at time instances $t^\ast=0.13\,\mathrm{s}$ and $t^\ast=0.17\,\mathrm{s}$ for Case NV2. Symbols correspond to the results obtained by \texttt{rheoMultiFluidInterFoam} solver or previous studies of Zainali et al. \cite{Zainali2013} and Vahabi and Sadeghy \cite{Vahabi2015}; Lines denote results yielded by the \texttt{rheoInterFoam} solver. (d) Time evolution of the droplet rise velocity for Case NV2. Lines: present work; Symbols: von Danwitz \cite{Danwitz2015}. Case NN: $Re = 35$, $We = 10$, $r_\rho = 0.1$, and $r_\mu = 0.1$. Case NV1: $Re = 1.419$, $We = 35$, $Wi = 1$, $\beta = 0.5238$, $r_\rho = 0.1$, and $r_\mu = 0.1$. Case NV2: $Re = 1.419$, $We = 35.28$, $Wi = 8.083$, $\beta_1 = 0.07$, $r_\rho = 0.1$, and $r_\mu = 0.1$. }
\label{Fig-viscoelastic}
\end{figure}

Figs. \ref{Fig-viscoelastic} (a,b) show a comparison of the results based on our new multiphase solver ($C_{a_i}=1,\,n_{lfi}=1$) with those of Hysing et al. \cite{Hysing2009} and Prieto \cite{Prieto2015} for a Newtonian (Case NN: $Wi=0$, $\beta_1=1$) and a viscoelastic surrounding fluid (Case NV: $Wi=1$, $\beta_1=0.5238$), respectively; for both cases the other parameters are $Re=35$, $We = 10$, $r_\rho = 0.1$, and $r_\mu = 0.1$. In particular, the steady-state drop shape is shown in Fig. \ref{Fig-viscoelastic}(a) and the time evolution of the rise velocity is depicted in Fig. \ref{Fig-viscoelastic}(b). A uniform computational mesh consisting of $160 \times 320$ control volumes was adopted as in the study of Prieto \cite{Prieto2015}. The results of Hysing et al. \cite{Hysing2009} correspond to the TP2D method based on a mesh consisting of $320 \times 640$ cells. A good agreement of the present predictions with the previously published results can be seen for both cases.

To further validate our code, we consider a second test case with $Re = 1.419$, $We = 35.28$, $Wi = 8.083$, $\beta_1 = 0.07$, $r_\rho = 0.1$, and $r_\mu = 0.1$ and compare the results of our new solver against the results of Zainali et al. \cite{Zainali2013} and Vahabi and Sadeghy \cite{Vahabi2015}. Following these works, a spherical droplet is initially placed in a 2D rectangle computational domain of width $L_x^\ast/L_{ch}^\ast=3.333$ and height $L_y^\ast/L_{ch}^\ast=6.666$, while its center is located at ($1/2\,L_x^\ast/L_{ch}^\ast$, $1/4\,L_y^\ast/L_{ch}^\ast$). No-slip conditions are enforced at all boundaries. For a droplet with radius $R_{d,0}^\ast=0.3\,\mathrm{cm}$ the dimensionless parameters correspond to a system with the following physical properties: $\rho_1^\ast=1000 \, \mathrm{kg}\, \mathrm{m}^{-3}$, $\rho_2^\ast=100 \, \mathrm{kg}\,\mathrm{m}^{-3}$, $\mu_{s,1}^\ast=0.0717 \, \mathrm{Pa} \, \mathrm{s}$, $\mu_{p,1}^\ast = 0.9533 \, \mathrm{Pa}\,\mathrm{s}$, $\mu_2^\ast = 0.1025 \, \mathrm{Pa}\, \mathrm{s}$, $\lambda_1^\ast =0.2\,\mathrm{s}$, and $g^\ast = 9.8\,\mathrm{m} \, \mathrm{s}^{-2}$. The grid resolution is $120\times240$ control volumes, while interface sharpening and filtering operations are considered with $C_{a_i}=1,\,n_{lfi}=1$. The computations of Zainali et al. \cite{Zainali2013} were performed by using an incompressible smoothed particle hydrodynamics method, while Vahabi et al. \cite{Vahabi2015} utilized a weakly-compressible smoothed particle hydrodynamics approach. The Oldroyd-B constitutive model was used in both works, and therefore for our computations we set $\epsilon_1=\zeta_1=0$.

Fig. \ref{Fig-viscoelastic}(c) shows a comparison for the droplet shape at $t^\ast = 0.13 \,\mathrm{s}$. The droplet shape predicted by the \texttt{rheoMultiFluidInterFoam} solver exhibits a cusped trailing edge which is a common feature of a Newtonian droplet in viscoelastic medium at high polymer concentrations \cite{Amirnia2013, Ohta2015}. We note that, for the same dimensional time instant, our prediction is somewhere in-between the sharper shape of Vahabi and Sadeghy \cite{Vahabi2015} and the smoother shape of Zainali et al. \cite{Zainali2013}. As shown in Fig. \ref{Fig-viscoelastic}(c), the cusp at the trailing edge becomes sharper at a later time ($t^\ast=0.17\,\mathrm{s}$) based on our multiphase solver. Even at that time instant, an equilibrium shape has not been established (a longer time is required $t^\ast>0.3\,s$ leading to an elongated cusp edge). The exact reasons for this discrepancy are not clear. To check whether this might be an effect of the grid resolution we have performed a mesh refinement study, which showed an negligible effect without influencing the above-mentioned observations. For instance, the maximum (minimum) $y^\ast$ position of the interface denoting the droplet shape (depicted by the $\alpha_2=0.5$ isoline) varied from $1.885\,\mathrm{cm}$ ($1.034\,\mathrm{cm}$) to $1.89\,\mathrm{cm}$ ($1.04\,\mathrm{cm}$) as the mesh was refined from $120\times240$ up to $320\times640$ control volumes. We have additionally performed simulations using the \texttt{rheoInterFoam} solver of the \texttt{RheoTool} toolbox \cite{rheoTool2016} to simulate this two-phase flow case, which produced almost identical results to those of our \texttt{rheoMultiFluidInterFoam} solver (see dashed lines in Fig. \ref{Fig-viscoelastic}(c)). Therefore, it may be reasonable to assume that these discrepancies could be attributed to the different numerical scheme employed by these authors. The latter conclusion is also supported by the comparison presented in Fig. \ref{Fig-viscoelastic}(d) where we provide a comparison with the predictions of von Danwitz \cite{Danwitz2015} for the exact same case. We note that von Danwitz \cite{Danwitz2015} employed a deforming-spatial-domain/stabilized space-time (DSD/SST) procedure that allows computations on moving meshes using the finite element method and reported similar differences with the results of Vahabi and Sadeghy \cite{Vahabi2015} and Zainali et al. \cite{Zainali2013}. He also provided results about the evolution of the rise velocity of the droplet and the direct comparison between our predictions and those by von Danwitz \cite{Danwitz2015} illustrates a good agreement as shown in Fig. \ref{Fig-viscoelastic}(d).

\subsubsection{Spreading of a droplet over a viscoplastic fluid} \label{sec3.2.3}

\begin{figure}[t!]
\centerline{\includegraphics[width=0.6\linewidth] {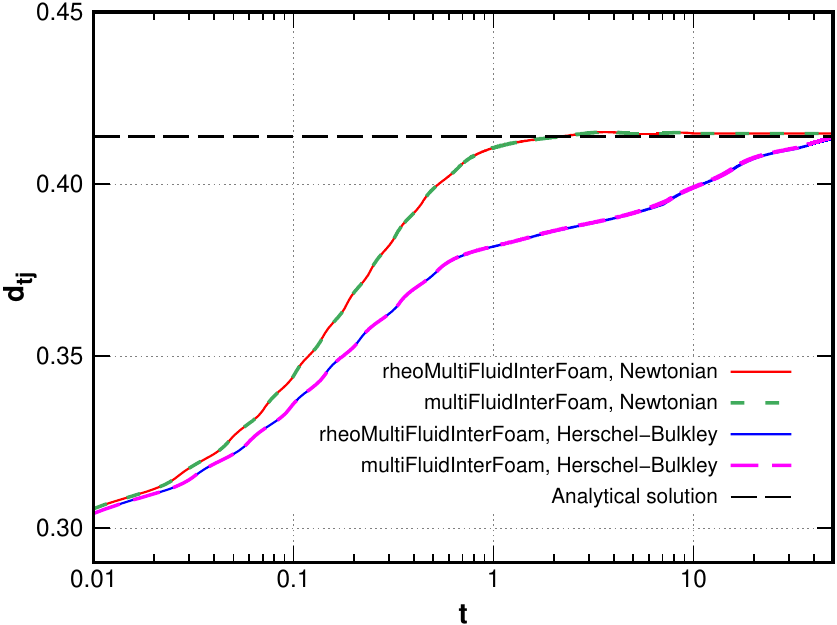}}
\caption{Time evolution of the interfaces of a floating lens with a non-Newtonian lower fluid for $Re = 1$ and $We_{p,1}=We_{p,2}=We_{p,3}=60$. The results are obtained by using the \texttt{multiFluidInterFoam} and \texttt{rheoMultiFluidInterFoam} solvers together with the bounded bi-viscosity Herschel--Bulkley model, $Bn = \tau_{0,1}^\ast R_d^\ast /(k_1^\ast U^\ast) = 0.15$, $\mu_{0,1}^\ast / k_1^\ast = 100$, and $n_1=1$.}
\label{figAA1}
\end{figure}

Now that we have verified the consistency of our solver for two-phase flows, we turn our attention to typical three-phase flows. Thus,in this Section we re-examine the floating lens problem, considering that the lower fluid layer is now a Herschel--Bulkley fluid. The interest is mostly focused on verifying whether the \texttt{rheoMultiFluidInterFoam} solver coupled with the bi-viscosity Herschel--Bulkley model available in the \texttt{RheoTool} toolbox produce the same results with the \texttt{multiFluidInterFoam} solver. This consistency check is verified in Fig. \ref{figAA1}, which shows negligible differences in the time evolution of $d_{tj}$ and the droplet shape obtained by the two new multiphase solvers. The Reynolds number is $Re=1$ and the phasic Weber numbers are $We_{p,1}=We_{p,2}=We_{p,3}=60$, whereas all the numerical details and parameters are exactly the same as those described in Section \ref{sec3.3.2}. The parameters related to the Herschel--Bulkley model are $Bn = \tau_{0,1}^\ast R_d^\ast /(k_1^\ast U^\ast) =0.15$, $\mu_{0,1}^\ast / k_1^\ast = 100$, and $n_1=1$.

\begin{figure}[h!]
\centerline{\includegraphics[width=0.85\linewidth] {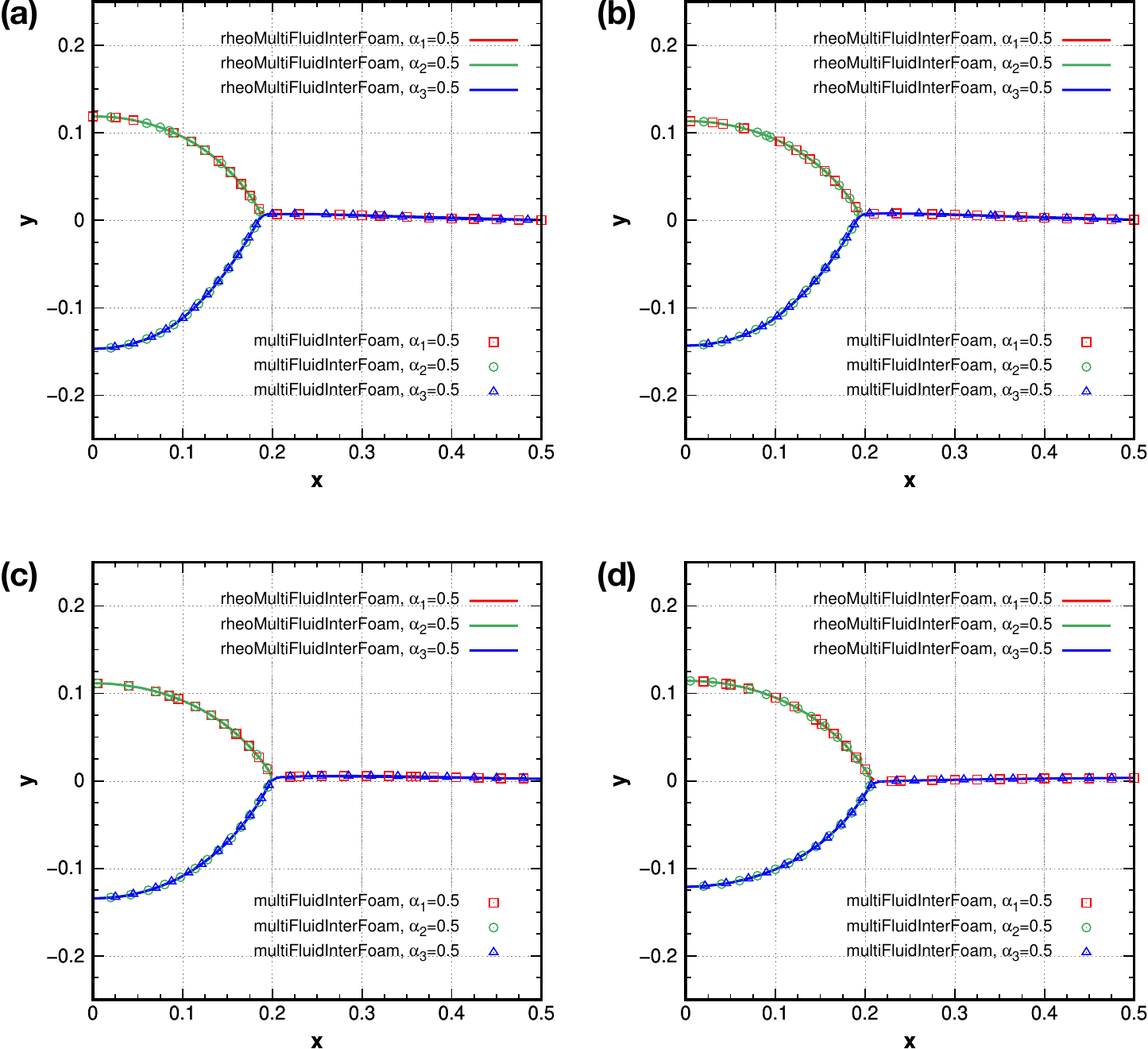}}
\caption{Shape of a floating lens with a non-Newtonian lower fluid for $Re=1$ and $We_{p,1}=We_{p,2}=We_{p,3}=60$ at non-dimensional times $t=1$ (a), $t=10$ (b), $t=25$ (c), and $t=50$ (d). The results are obtained by using the \texttt{multiFluidInterFoam} and \texttt{rheoMultiFluidInterFoam} solvers together with the bounded bi-viscosity Herschel--Bulkley model, $Bn = \tau_{0,1}^\ast R_d^\ast /(k_1^\ast U^\ast) = 0.15$, $\mu_{0,1}^\ast / k_1^\ast = 100$, and $n_1=1$.}
\label{figAA2}
\end{figure}

Fig. \ref{figAA1} indicates that a longer time is required in order to reach asymptotically the steady-state solution in the presence of the lower non-Newtonian fluid. Two regions are evident; the first region can be identified at early times, where there is a rapid increase of the droplet's length $d_{tj}$ with time and the second one where a slow increase of $d_{tj}$ takes place. As the droplet is initially away from a stable regime, it is suddenly put in motion. The stress in the region adjacent to the droplet at the non-Newtonian fluid is greater than the yield stress and it eventually starts to flow. In this region, because of the intense shear stresses the effective viscosity of the subphase decreases considerably and becomes much smaller than that of the fluid away from the contact line. Consequently, the material yields only adjacent to the droplet, whereas the rest remains unyielded with a large effective viscosity, $\mu_{0,1}^\ast$. As a result, we find that the spreading of the droplet resembles that of the pure Newtonian case, as suggested by the similar slopes in the $d_{tj}$ curves at early times. The lower fluid and the droplet are decelerated and the magnitude of the velocities is reduced, delaying the spreading process. As the spreading proceeds and the droplet approaches its equilibrium state, the level of stresses experienced in the subphase decreases considerably and the size of the yielded region adjacent of the droplet decreases with time. Eventually, at very late stages it is expected that when the experienced stress in the bottom fluid becomes smaller than the yield stress and the whole lower fluid may become unyielded, impeding droplet spreading. However, we have to note that the use of a regularized model does not allow to accurately predict the droplet entrapment by unyield material, since the material below yield stress is effectively described by a fluid with very large viscosity. Therefore, the regularized model allows for creeping flow motion of the viscoplastic material and the droplet eventually reaches its equilibrium shape. Since the steady-state solution is independent of the viscosities of the phases, and the predicted geometrical features of the droplet at equilibrium shown in Fig. \ref{figAA2} are found to be the same as in the Newtonian case described in Section \ref{sec3.3}.

\subsubsection{Levitation of a drop within Newtonian/viscoplastic fluids}

\begin{figure}[h!]
\centerline{\includegraphics[width=1.0\linewidth] {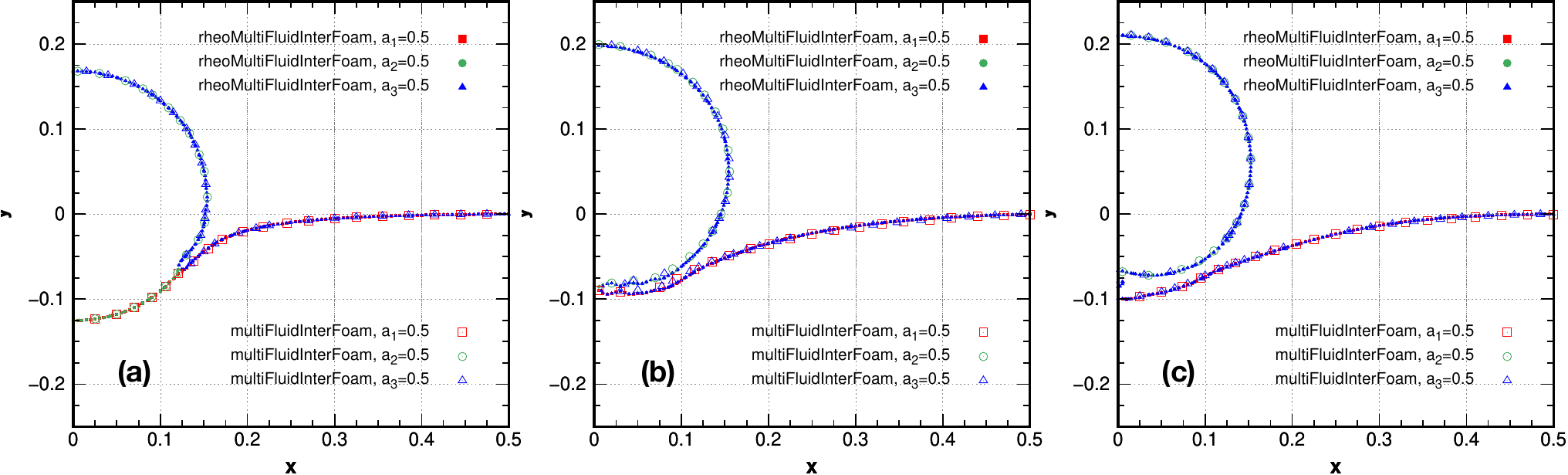}}
\caption{Comparison of the $a_1=a_2=a_3=0.5$ level contour for all phases at non-dimensional times $t=0.01$ (a), $t=0.1$, (b), and $t=1$ (c) in the case of levitating drop with a non-Newtonian lower fluid. Results are obtained based on the \texttt{multiFluidInterFoam} and \texttt{rheoMultiFluidInterFoam} solvers with the bounded bi-viscosity Herschel--Bulkley model, $Bn = \tau_{0,1}^\ast R_d^\ast /(k_1^\ast U^\ast) = 0.15$, $\mu_{0,1}^\ast / k_1^\ast = 100$, and $n_1=1$. The relevant parameters are the same to those for the case of Fig. \ref{figC1}.}
\label{figC2}
\end{figure}

In this Section, the simulations for the drop levitation are repeated with the lower fluid exhibiting a viscoplastic behavior. Again, our interest is mostly focused on verifying whether the two new multiphase solvers coupled with a bi-viscosity Herschel--Bulkley model produce the same results. This consistency check is verified in Fig. \ref{figC2}, which shows negligible differences in the time evolution of the droplet shape and interfaces obtained by the new solvers. This is expected since the only difference between the two solvers is that the constitutive models of \texttt{RheoTool} toolbox are used the \texttt{rheoMultiFluidInterFoam} solver, and comparisons are performed for the same viscosity model. All the numerical details and other parameters are exactly the same as those described above for the Newtonian case. The parameters related to the Herschel--Bulkley model are $Bn = \tau_{0,1}^\ast R_d^\ast /(k_1^\ast U^\ast) = 0.15$, $\mu_{0,1}^\ast / k_1^\ast = 100$, and $n_1=1$.

\section{Discussion}\label{sec4}

The most significant difference between \texttt{multiFluidInterFoam}, \texttt{rheoMultiFluidInterFoam} solvers and the original \texttt{multiphaseInterFoam} solver is the treatment of the continuum surface tension force. It can be shown that both Eq. (\ref{eqA11}) of the original solver and Eq. (\ref{eqB4}) of the new solvers produce the correct expression for the surface tension force in case of two-phase flows. However, even though the two CSF models are identical theoretically, their numerical behaviour differs significantly. The CSF model used in the original solver involves the calculation of more gradients of the phase fraction, which may introduce additional numerical errors, especially for highly discontinuous $a_i$ fields. These errors, though, can be mitigated by smoothing the distribution of $a_i$ in the calculation of curvature. 

In the dambreak problem, the flow is mainly controlled by differences in the density of the phases. The interfacial tension effects are secondary and they do not contribute considerably to the dynamics of the multiphase flow. Any three-phase regions, such as contact points, disappear during the early stages of the flow evolution and mostly interfaces between two phases are observed. Therefore, for this problem, the particular treatment of the continuum surface tension force does not play a significant role; similar results are obtained between the standard \texttt{multiphaseInterfoam} and both our newly developed solvers with only some small differences at the very late stages of the simulation, which can clearly be attributed to the different numerical implementation of the CSF model.

In the case of interfacial tension dominated multiphase flows, though, it is demonstrated that the original \texttt{multiphaseInterfoam} solver is not able to produce the correct behavior, providing inaccurate predictions. The obtained results contradict both theory and previously published results for both problems of a floating lens and a levitating drop. On the other hand, we show that the newly developed \texttt{multiFluidInterFoam} solver and its non-Newtonian counterpart (\texttt{multiFluidInterFoam}) improve significantly the quality of the predictions, exhibiting robust behaviour and providing very accurate predictions, as indicated by the good agreement against analytical and previous numerical results.

Finally, we have shown that the use of the \texttt{PIMPLE} velocity-pressure algorithm turned out to be insufficient in cases involving a non-Newtonian fluid. For example, in the floating lens and the rising bubble problems with a Herschel--Bulkley fluid, although the numerical solutions do not necessarily diverge, the results were found to be quite sensitive to the choice of the time step. In some case, the interactions between a very viscous fluid and the wall induced numerical instabilities that diffused in the inner flow region. In this way, we often found that the developed shear stress was greater than the yield stress and the whole lower fluid layer was put in motion eliminating the presence any regions with unyielded material. Moreover, we noticed several regions of local maximum values of viscosity corresponding to the unyielded region appear randomly in the non-Newtonian fluid, with the droplet or the bubble mostly located adjacent to a fluid having a lower viscosity. This resulted in nonphysical unsteady solutions for the rising bubble problem. In addition, the transient dynamics of the droplet spreading were influenced, indicating an unexpected and nonphysical fast evolution of the spreading process. In order to obtain a reasonable and accurate solution, we found that the time step must be reduced or the inner iterations of the \texttt{PIMPLE} algorithm must be increased. However, both choices increase considerably the overall computational cost. The adoption of the \texttt{SIMPLEC} algorithm allows to use higher time steps without sacrificing the accuracy of the solutions, obtaining also reliable results. This is particularly important, since complex fluid flows are typically more computationally demanding and e.g. in the case of droplet spreading, the non-Newtonian case generally requires greater times to reach a steady-state regime.

\section{Conclusions}\label{sec5}

In this article, a robust and accurate finite-volume solver \texttt{multiFluidInterFoam} for the simulation of three-phase flows of Newtonian/non-Newtonian fluids with significant surface tension effects is developed in the open source \texttt{OpenFOAM}. The volume of fluid method (VOF) is utilized, which employs a volumetric phase fraction for each fluid phase. Sharpening of the interface is achieved by introducing an artificial compression term in the equation for each volume fraction. Starting from the original \texttt{multiphaseInterFoam} solver, we implemented the interfacial tension coefficient decomposition method to deal with tension pairings between different phases using a compositional approach. The resulting interfacial force is reformulated properly as the sum of each component body force. VOF smoothing is also performed to estimate accurately the curvature of each interface. The \texttt{SIMPLEC} algorithm was adopted to utilize the velocity-pressure coupling. 

Finally, a second solver \texttt{rheoMultiFluidInterFoam} was developed by implementing the \texttt{RheoTool} toolbox in order to make use of several constitutive models for complex fluids. This allows the investigation of more physically and rheologically complex situations in which several non-Newtonian properties such as viscoelasticity, shear-thinning, and yield stress are taken into account. 

The numerical framework has been validated against the three-phase dam break problem and several benchmark problems for a floating lens and a levitating droplet for Newtonian fluids. The performance of the \texttt{rheoMultiFluidInterFoam} solver was assessed in the case of two-phase rising bubble in a viscoplastic or a viscoelastic fluid and several consistency tests were performed for the floating lens and levitating drop, in which the lower fluid exhibited a viscoplastic behavior. The newly developed multiphase solvers can represent better the underlying physics of surface tension dominated flow problems and decrease the overall computational cost without sacrificing accuracy. 
 
\section*{Acknowledgments}
This project has received funding from the Hellenic Foundation for Research and Innovation (HFRI) and the General Secretariat for Research and Technology (GSRT), under grant agreement No 792. Results presented in this work have been produced using the Aristotle University of Thessaloniki (AUTh) High Performance Computing Infrastructure and Resources.

\subsection*{Appendix: Derivation of VOF sharpening for multiphase flows}
%
%
A mixed velocity $\mathbf{u}^\ast$ can be defined as a weighted average of the component velocities $\mathbf{u}_k^\ast$ as follows
\begin{equation}
\mathbf{u}^\ast=\sum_k a_k \mathbf{u}^\ast_k.
\label{AVOFComp2}
\end{equation}

\noindent
The relative velocity between fluids $i$ and $k$ can be defined as
\begin{equation}
\mathbf{u_r}^\ast_{,ik}=\mathbf{u}^\ast_i-\mathbf{u}^\ast_k.
\label{AVOFComp5}
\end{equation}

\noindent
The mixed velocity in Eq. (\ref{AVOFComp2}) can be re-written as 
\begin{equation}
\mathbf{u}^\ast=\sum_{k \neq i} a_k \mathbf{u}^\ast_k +  a_i \mathbf{u}^\ast_i \Rightarrow a_i \mathbf{u}^\ast_i = \mathbf{u}^\ast - \sum_{k \neq i} a_k \mathbf{u}^\ast_k.
\label{AVOFComp4}
\end{equation}

\noindent
Multiplying Eq. (\ref{AVOFComp4}) by $a_i$, it yields
\begin{equation}
a_i a_i \mathbf{u}^\ast_i 
=  a_i \mathbf{u}^\ast - a_i \sum_{k \neq i} a_k \mathbf{u}^\ast_k. \label{AVOFComp8a}
\end{equation}

\noindent
The term $a_k\mathbf{u}^\ast_k$ in the above equation can be rearranged further, by solving Eq. (\ref{AVOFComp5}) to $\mathbf{u}_k$ and multiplying the result by $a_k$, and Eq. (\ref{AVOFComp8a}) can be written as  
\begin{equation}
a_i a_i \mathbf{u}^\ast_i =  a_i \mathbf{u}^\ast - a_i \sum_{k \neq i} \left( a_k \mathbf{u}^\ast_i - a_k \mathbf{u_r}_{,ik} \right). \label{AVOFComp8}
\end{equation}

\noindent
After some maths, the following is obtained
\begin{equation}
a_i \mathbf{u}^\ast_i = a_i \mathbf{u}^\ast + a_i \sum_{k \neq i}  a_k \mathbf{u_r}^\ast_{,ik}.
\label{AVOFComp9}
\end{equation}

\noindent
Finally, Eq. (\ref{VOFComp1}) that is numerically solved in the multiphase solvers can be derived from the transport equation (Eq. \ref{eqA4}) for the $i$th volume fraction $a_i$ by replacing the term $a_i \mathbf{u}^\ast_i$ with the above relation (\ref{AVOFComp9}).

\section*{References}


\end{document}